\documentstyle[11pt]{article}
\def\ut#1{\rlap{\lower1ex\hbox{$\sim$}}#1{}}
\begin{document}
\begin{flushright}
\renewcommand{\textfraction}{0}
\baselineskip=15pt
UU-REL-93/1/9  \\
hep-th/9301028 \\
January 9th 1992\\
\end{flushright}
\vskip 3cm
\renewcommand{\thefootnote}{\fnsymbol{footnote}}
\begin{center}
{\LARGE {\bf Knot Theory and Quantum Gravity \\
in Loop Space: A Primer
\footnote{To appear in ``Proceedings of the Vth Mexican School of
Particles and Fields'', J. L. Lucio, editor, World Scientific, Singapore
(1993).}
}}
\end{center}
\renewcommand{\thefootnote}{\arabic{footnote}}
\setcounter{footnote}{0}
\begin{center}
Jorge Pullin \\
{\it Department of Physics,  University of Utah} \\
{\it Salt Lake City, UT 84112 USA}
\end{center}

\begin{abstract}
These notes summarize the lectures delivered in the V Mexican School
of Particle Physics, at the University of Guanajuato. We give a
survey of the application of Ashtekar's variables to the quantization
of General Relativity in four dimensions with special emphasis on the
application of techniques of analytic knot theory to the loop
representation.  We discuss the role that the Jones Polynomial plays
as a generator of nondegenerate quantum states of the gravitational
field.
\end{abstract}

\section{Quantum Gravity: why and how?}

I wish to thank the organizers for inviting me to speak here. This may
well be a sign of our times, that a person generally perceived as a
``General Relativist'' would be invited to speak at a Particle Physics
School. It just reflects the higher degree of interplay these two
fields have enjoyed over the last years. In these lectures we will see
more reasons for this enhanced interplay. We will see several notions
from Gauge Theories, as Wilson Loops for instance, playing a central
role in gravitation. An even greater interplay takes place with
Topological Field Theories. We will see the important role that the
Chern-Simons form, the Jones Polynomial and other notions of knot
theory seem to play in General Relativity.

The quantization of General Relativity is a problem that has defied
resolution for the last sixty years. In spite of the long time that
has been invested in trying to solve it, we believe that several
people do not necessarily fully appreciate the reasons of our failure
and the magnitude of the problem. It is a general perception
--especially among particle physicists-- that ``General Relativity is
nonrenormalizable'' and that is the basic problem with the theory.
This statement is misleading in three ways:

a) The fact that a theory is nonrenormalizable does not necessarily
mean that the theory has an intrinsic problem or is ``bad'' in any way.
It merely says that perturbation theory does not apply to the problem
in question. As we will see in c) there are actually good reasons to
believe that ordinary perturbation theory {\em should} fail for
General Relativity.

b) Deciding if a theory is or is not renormalizable can be quite
tricky.  The prime example is 2+1 dimensional gravity, which most
people thought to share the renormalizability pathologies of 3+1
gravity until Witten \cite{Wi2+1} pointed out that it could be exactly
solved. A posteriori it was of course found that the theory is in fact
renormalizable \cite{De}.

c) There actually are very good reasons why we should expect General
Relativity to have ``problems'' (we would rather call them
``subtleties'' or ``challenges'') in quantization. Prime among them is
the issue of diffeomorphism invariance, which in turn implies other
problems as the lack of observables for the theory. This last problem
clearly reflects the nature of the issue: even if we were somehow able
to make General Relativity renormalizable, we would not know {\em
what} to compute with such a perturbatively well behaved theory. Even
if we were able to compute something we would not know how to interpret
it.

Due to these and other arguments, we believe that perturbative
quantization of General Relativity may well be a red herring. So too
may also be the idea of abandoning General Relativity in favour of other
theories that present some particular better behaviour (usually only
apparent) when perturbatively quantized. We repeat: even if we had a
perfectly renormalizable theory of quantum gravity, it is little what
we could actually {\em do} with it until we address the fundamental
questions of what kind of physics can we do in a diffeomorphism
invariant context.

If one is interested in these kind of questions, in particular the
issue of diffeomorphism invariance, nonperturbative
quantization seems the way to go. There are several options if one
wants to attempt a nonperturbative quantization of gravity, ranging
from quite radical to very conservative ones. Of all these, probably
the most conservative is Canonical Quantization. After all,
this was the first method of quantization ever invented and is the one
most physicists feel comfortable with. Canonical Quantization
therefore seems an attractive approach to study the issues that arise
in the quantization of General Relativity. If the resulting theory
makes sense, one expects that other quantization techniques would in
the end give the same results.

We will discuss in these lectures the Canonical Quantization of
General Relativity in four dimensions. As we have argued, the use of
other theories or number of dimensions seems at the moment
superfluous. We do not even understand --for good reasons-- arguably
the simplest theory (General Relativity). Therefore it seems to make
little sense for our purposes to embark on the study of more
complicated theories. It may well be that at some point it becomes
apparent that General Relativity does not furnish a suitable base for
a theory of Quantum Gravity. Until that point is reached we think it
is useful as the simplest theory of gravity that has all the desired
features one would expect in such a theory.

We will therefore proceed with a very conservative Canonical
Quantization scheme but we will pursue a slight variant from the
traditional approaches. We will use a new set of canonical variables
for the treatment of Hamiltonian General Relativity, the Ashtekar
Variables \cite{As}.
These variables have the advantage of casting General
Relativity in a fashion that closely resembles Yang-Mills theories.
This will allow us to introduce several useful techniques from Yang-Mills
theories into General Relativity.

The plan of these lectures is as follows: in section 2 we discuss the
traditional canonical formulation of General Relativity. In section 3
we introduce the Ashtekar New Variables and discuss the classical
theory.  In section 4 we discuss the quantum theory in the connection
representation, and point out the role that Wilson Loops and the
Chern-Simons form play in the theory. In section 5 we present the Loop
Representation and develop technology for dealing with the constraints
and the wavefunctions written in terms of loops. In section 6 we
discuss various aspects of Knot Theory.  In section 7 we make use of
the results of sections 5 and 6 to construct a family of nondegenerate
physical states of Quantum Gravity in terms of knot invariants. We end
in section 8 with some final remarks and a general discussion of the
present status of the program.

\section{Brief Summary of General Relativity and Canonical
Quantization}
\subsection{Classical General Relativity}
General Relativity is a theory of gravity in which the gravitational
interaction is accounted for by a deformation of spacetime. The
fundamental variable for the theory is the spacetime metric
$g_{ab}$. The action for the theory is given by,

\begin{equation}
S = \int d^{4}x \sqrt{-g} R(g_{ab}) +\int d^{4}x \sqrt{-g}
{\cal L}({\rm matter})
\end{equation}
where $g$ is the determinant of $g_{ab}$, $R(g_{ab})$ is the
curvature scalar and we have included also a term to take into account
possible couplings to matter, although in these lectures we will
largely concentrate on the theory in vacuum. The equations of motion
for this action, obtained by varying the action with respect to
$g_{ab}$ are,
\begin{equation}
R_{ab}-{\textstyle {1 \over 2}} g_{ab} R = {\delta S_{\rm
matter}\over \delta g^{ab}}
\end{equation}
and are called Einstein Equations. The theory is invariant under
diffeomorphisms on the four manifold (coordinate transformations). This
means it has a symmetry. The Einstein Equations are in principle ten
equations (all the tensors are symmetric and therefore only have ten
independent components). However, due to the presence of the
diffeomorphism symmetry, several of the equations are redundant. This
issue is best seen in the canonical formalism. We will briefly set it up
in the next two subsections. The reader wanting a more detailed treatment
can find it in references \cite{Asbi,Asws}.

\subsection{Canonical formulation, the 3+1 split}
To cast General Relativity in a canonical form, we need to split
space-time into space and time. Without a notion of time, we do not
have a notion of evolution, and therefore no notion of
``Hamiltonian''. This may seem strange at first. One of the great
accomplishments of Relativity was to put space and time into an equal
footing, and now we seem to be destroying it. We will show that this
is not the case. Although the Canonical formalism manifestly breaks
the covariance of the theory by singling out a particular time
direction, in the end the formalism tells us that it really did not
matter which time direction we took. That is, the covariance is recovered
implicitly in the theory, and the time picked is only a ``fiducial''
one for constructional purposes. We will see more details of this
fairly soon.

\begin{figure}
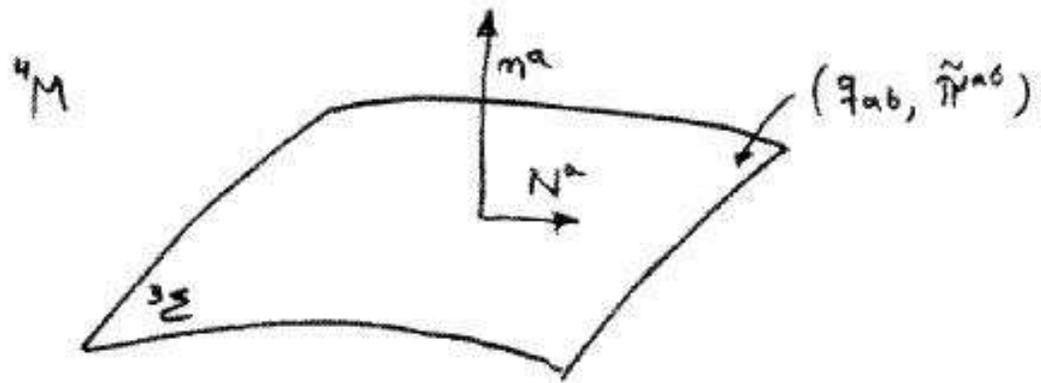

\vskip 6cm
\caption{3+1 foliation of spacetime and variables of the canonical formalism}
\label{foliation}
\end{figure}

So we foliate spacetime ${}^{4}M$ into a spatial manifold
${}^{3}\Sigma$ and a time direction $t^{a}$, as shown in figure
\ref{foliation}. We now decompose the
time direction into components
normal and perpendicular to the three surface,
\begin{equation}
t^{a}=N n^{a}+ N^{a}
\end{equation}
where $n^{a}$ is the normal to the three surfaces and $N^{a}$ is
tangent to the three surfaces, and is called the shift vector. The
scalar $N$ is called the lapse function. Given the metric $g_{ab}$
and the timelike vector $n^{a}$ one can define a positive-definite
(Euclidean) metric on the three surface $q^{ab}=g^{ab}+n^{a}n^{b}$.
Another important quantity on the three surface is its extrinsic
curvature, defined by $K_{ab}=q_{a}^{m} q_{b}^{n} \nabla_{m} n_{n}$,
where $\nabla$ is the covariant derivative compatible with $g_{ab}$.
To clarify the role of the extrinsic curvature it is enough to compute
the ``time derivative'' of the three metric, given by its Lie
derivative with respect to $t^{a}$ (exercise),
\begin{equation}
\dot{q}_{ab}={\cal L}_{\vec{t}}\ q_{ab} = 2 N K_{ab}+{\cal L}_{\vec{N}} q_{ab}
\end{equation}
So we see that the role of the extrinsic curvature is roughly that of
``time derivative'' of the three-metric, giving an idea of how the
three dimensional surface is deformed with respect to the ambient four
dimensional spacetime. One can rewrite the action of General
Relativity in terms of these quantities in the following form,
\begin{eqnarray}
S &=& \int dt\ L(q,\dot{q}) \\
L(q,\dot{q}) &=& \int d^{3}x N \sqrt{q} ({}^{3}R+K_{ab}K^{ab}-K^{2})
\end{eqnarray}
where t is a parameter along the integral curves of the vector
$t^{a}$. ${}^{3}R$ is the scalar curvature of the three dimensional
surface. To achieve this form of the action one needs to neglect
surface terms. Along these lectures we will always assume that the
spatial three-surface is compact, as in the case of some cosmologies.
One could treat the asymptotically-flat case (which includes for
instance, stars and black-holes) by imposing appropriate boundary
conditions at infinity. This can be done in a reasonable
straightforward manner, although we will not discuss it here for
the sake of brevity.

We now have the action of General Relativity in a reasonable form to
formulate a canonical analysis. We have expressed it in terms of
variables that are functions of ``space'' (functions of the
three-surface) and that ``evolve in time''. This is the usual setup
for doing canonical formulations.

We pick as a canonical variable the three metric $q^{ab}$ and compute
its conjugate momentum,
\begin{equation}
\tilde{\pi}_{ab}=
{\delta L \over \delta \dot{q}^{ab}}=\sqrt{q} (K_{ab}-K q_{ab})
\end{equation}
(throughout these lectures we will denote tensor densities of weight +1
with a tilde and those of weight -1 with an undertilde).
We see that the ``conjugate momentum'' to $q_{ab}$ is roughly related
to the extrinsic curvature (``time derivative'').

We are now in position to perform the Legendre transform and obtain
the Hamiltonian of the theory,
\begin{equation}
H(\pi,q) = \int d^{3}x \, (\tilde{\pi}_{ab}\dot{q}^{ab} -L)
\end{equation}
and replacing $\dot{q}$ in terms of $\tilde{\pi}$, we get
(exercise),
\begin{equation}
H(\pi,q) = \int d^{3}x\,( N (-q^{1/2} R +q^{-1/2}(\tilde{\pi}^{ab}
\tilde{\pi}_{ab}-{\textstyle {1 \over 2}} \tilde{\tilde{\pi}}^{2}) )
-2  N^{b} D_{a} \tilde{\pi}^{a}_{b})
\end{equation}
where $D_{a}$ is the covariant derivative on the three surface
compatible with $q_{ab}$.

\subsection{The constraints}
Having done this, let us step back a minute and analyze the formalism
we built. We started from a four dimensional metric $g^{ab}$ and we
now have in its place the three dimensional $q^{ab}$ and the ``lapse''
and ``shift'' functions $N$ and $N^{a}$. We defined a conjugate
momentum for $q_{ab}$. However, notice that nowhere in the formalism
does a time derivative of the lapse or shift appear. That means their
conjugate momenta are zero. That is, our theory has constraints. In
fact, if we rewrite the action using the expression for the
Hamiltonian we give above, we get,
\begin{equation}
S = \int dt \int d^{3}x\,( (\tilde{\pi}_{ab}\dot{q}^{ab} +
\ut{N} (-q R +(\tilde{\pi}^{ab}
\tilde{\pi}_{ab}-{\textstyle {1 \over 2}}
\tilde{\tilde{\pi}}^{2})) -2  N^{b} D_{a}
\tilde{\pi}^{a}_{b})
\end{equation}
and if we vary it with respect to $\ut{N}$ and $N^{b}$ in order to get
their respective equations of motion, we get four expressions,
functions of $\tilde{\pi}$ and $q$ which should vanish identically, and are
usually called $\tilde{C}^{a}$ and $\tilde{\tilde{C}}$,
\begin{eqnarray}
\tilde{C}_{a}(\pi,q) &=& 2   D_{b} \tilde{\pi}^{b}_{a} \label{constraints1}\\
\tilde{\tilde{C}}(\pi,q) &=& -\tilde{\tilde{q}} R +(\tilde{\pi}^{ab}
\tilde{\pi}_{ab}-{\textstyle {1 \over 2}} \tilde{\pi}^{2})\label{constraints2}
\end{eqnarray}
For calculational simplicity, these equations are usually  ``smoothed out''
with arbitrary test fields on the three manifold,
$C(\vec{N})=\int d^{3}x N^{a} \tilde{C}_{a}$, $C(\ut{N})=\int d^{3}x \ut{N}
\tilde{\tilde{C}}$. (Notice
that the notation is unambiguous. One can write the constraints now as
$C(\ut{N})=0$ and $C(\vec{N})=0$, due to the arbitrariness of the test
fields).

Notice that these equations are ``instantaneous'' laws, i.e.\ they
must be satisfied {\em on each hypersurface}. They tell us that if we
want to prescribe data for a gravitational field, not every pair of
$\tilde{\pi}$ and $q$ will do, eqs. (\ref{constraints1},
\ref{constraints2}) should be satisfied.
(Notice that there are six degrees of freedom in $q^{ab}$, subjected
to four constraints, this leaves us with two degrees of freedom for
the gravitational field, as expected).

These equations have the same character as the Gauss Law has for
electromagnetism, which tells us that any vector field would not
necessarily work as an
electric field, it must have vanishing divergence in vacuum. As
is well known, the Gauss Law appears as a consequence of the $U(1)$
invariance of the Maxwell equations. An analogous situation appears
here. To understand this, consider the Poisson bracket of any quantity
with the constraint $C(\vec{N})$. It is easy to check that (exercise),
\begin{equation}
\{ f(\tilde{\pi},q), C(\vec{N})\} = {\cal L}_{\vec{N}} f(\tilde{\pi},q).
\end{equation}
Therefore we see that the constraint $C(\vec{N})$ ``Lie drags'' the
function $f(\pi,q)$ along the vector $\vec{N}$. Technically, it is the
infinitesimal generator of diffeomorphisms of the three manifold in
phases space. As
the Gauss law (in the canonical formulation of Maxwell's theory) is
the infinitesimal generator of $U(1)$ gauge transformations, the
constraint here is the infinitesimal generator of spatial
diffeomorphisms. This clearly shows why we have this constraint in the
theory: it is the canonical representation of the fact that the theory
is invariant under spatial diffeomorphisms. An analogous situation
stands for the  constraint $C(\ut{N})$, although we will not discuss it
in detail for reasons of space: it is the generator of ``time''
diffeomorphisms, and it is usually called the Hamiltonian constraint.

We can now work out the equations of motion of the theory either
varying the action with respect to $q^{ab}$ and $\tilde{\pi}_{ab}$ or taking
the Poisson bracket of these quantities with the Hamiltonian
constraint.

\subsection{Quantization}

Having set up the theory in canonical form we can now proceed and
attempt a Canonical Quantization. The process can be roughly
summarized in the following  sequence of steps. The reader may notice
that at each step there are many possible choices. Differente choices
will yield different quantizations.
\begin{enumerate}
  \item Pick up a complete set of canonical quantities that form an
	algebra under Poisson brackets. In many systems one simply
	takes $p$ and $q$, but one can pick other ``noncanonical''
	quantities. We will actually do so in the next chapters.
  \item Represent these quantities as operators acting on a space of
	(wave)functionals of ``half'' of the elements of the algebra.
	These operators will in turn give a ``quantum'' representation of the
	algebra defined in the previous point.  At this point the space of
	functions is not a Hilbert space.  One usual choice is to take
	functionals of $q$, $\Psi(q)$, and represent $\hat{q}$ as a
	multiplicative operator and $\hat{p}$ as a functional
	derivative, $\hat{p} \Psi(q) = {\delta \over \delta q} \Psi(q)$.
  \item One wants the wavefunctions to be invariant under the
	symmetries of the theory. As we saw the symmetries are
	represented in this language as constraints. The requirement
	that the wavefunctions be annihilated by the constraints
	(promoted to operatorial equations) implements this
	requirement. Not all functionals chosen in the previous steps
	will be annihilated by the quantum constraint equations. We
	call those who are ``physical states'' (notice that we still
	do not have a Hilbert space).
  \item One introduces an inner product on the space of physical
	states in order to compute expectation values and make
	physical predictions. Only at this point does one have an
	actual Hilbert Space. How to find this inner product is not
	prescribed by standard canonical quantization (we will discuss
	this in the next section).
	Under this inner product the physical states should be normalizable.
	The expectation values, by the way,
	only make sense for quantities that are invariant under the
	symmetries of the theory (quantities that classically have
	vanishing Poisson brackets with all the constraints). We call
	them physical observables. For the gravitational case {\em
	none} is known for compact spacetimes
	(we will return to this issue later). The observables
	of the theory should be self adjoint operators with respect to the
	inner product in order to yield real expectation values.
\end{enumerate}

Let us start by step one. As it is said, one can pick various choices
of algebras. Let us concentrate on the simplest one, just picking the
canonical variables we introduced $q^{ab}$ and $\tilde{\pi}_{ab}$. Their
Poisson bracket is, being canonically conjugate variables,
$\{q^{ab}(x), \tilde{\pi}_{cd}(y)\}= \delta(x-y) \delta^{a}_{c}
\delta^{b}_{d}$. Step 2 can be fulfilled taking wavefunctions
$\Psi(q^{ab})$ and representing $q^{ab}$ as a multiplicative operator
and $\tilde{\pi}_{ab}$ as a functional derivative.

It is in step 3 that we run into trouble. We have to promote the
constraints we discussed in the last subsection to quantum operators.
This in itself is a troublesome issue, since being General Relativity
a field theory,
issues of regularization and factor ordering appear. One can, ---at
least formally--- find factor orderings in which the diffeomorphism
constraint becomes the infinitesimal generator of diffeomorphisms on
the wavefunctions. Therefore the requirement that a wavefunction be
annihilated by it just translates itself in the fact that the
wavefunction has to be invariant under diffeomorphisms. This is not
difficult to accomplish (formally!). One simply requires that the
wavefunctions not actually be functionals of the three metric
$q^{ab}$, but of the ``three geometry'' (by this meaning the
properties of the three geometry invariant under diffeomorphisms).
That is, what we are saying is just
a restatement of the fact that the functional should be invariant
under diffeomorphisms. One can come up with several examples of
functionals that meet this requirement. The real trouble appears when
we want to make the wavefunctions annihilated by the Hamiltonian
constraint. This constraint does not have a simple geometrical
interpretation in terms of three dimensional quantities (remember that
the idea that it represent ``diffeomorphisms in time'' does not
help here, since we are always talking about equations that hold
{\em on the three surface} without any explicit reference to time).
Therefore we are just forced to proceed crudely: promote the
constraint to a wave equation, use some factor ordering (hopefully
with some physical motivation), pick some regularization and try to
solve the resulting equation. It turns out that this task was never
accomplished in general (it was in simplified minisuperspace examples).
Among the difficulties that conspire in this direction
is the fact that the constraint is a nonpolynomial function of the
basic variables (remember it involves the scalar curvature, a
nonpolynomial function of the three-metric).

Therefore the program of canonical quantization stalls here. Having
been unable to find the physical states of the theory we are in a bad
position to introduce an inner product (since we do not know on what
space of functionals to act) and actually make physical predictions.
This issue is compounded by the fact that we do not know any
observables for the system, which puts us in a more clueless situation
with respect to the inner product. This status of affairs was reached
already in the work of DeWitt in the 60's \cite{DeWi} and little
improvement was made until recently.
We will see in the next chapter
that the use of a new set of variables improves the situation with
respect to the Hamiltonian constraint, giving hopes of maybe allowing
us to attack the problem of the inner product.

\section{The Ashtekar New Variables}
The following three
subsections follow closely the treatment of \cite{Asws}. The reader is
referred to it for more detailed explanations.
\subsection{Tetradic General Relativity}
To introduce the New Variables, we first need to introduce the
notion of tetrads. In a nutshell, a tetrad is a vector basis in
terms of which the metric of spacetime looks locally flat. Mathematically,
\begin{equation}\label{tetrad}
g_{ab} = e_{a}^{I} e_{b}^{J} \eta_{IJ}
\end{equation}
where $\eta_{IJ}={\rm diag}(-1,1,1,1)$ is the Minkowski metric, and
equation (\ref{tetrad}) simply expresses that $g_{ab}$, when written
in terms of the basis $e_{a}^{I}$, is locally flat. If spacetime were
truly flat, one
could perform such a transformation globally, integrating the basis
vectors into a coordinate transformation $e_{a}^{I}={\partial x^{I}
\over \partial x'^{a}}$. In a curved spacetime these equations cannot
be integrated and the transformation to a flat space only works
locally, the flat space in question being the ``tangent space''. From
equation (\ref{tetrad}) it is immediate to see that given a tetrad,
one can  reconstruct the metric of spacetime. One can also
see that although $g_{ab}$ has only ten independent components,
the $e_{a}^{I}$ have sixteen. This is due to the fact that eq.
(\ref{tetrad}) is invariant under Lorentz transformations on the
indices $I,J\ldots{}$. That is, these indices behave as if living in
flat space. In summary, tetrads have all the information needed to
reconstruct the metric of spacetime but there are extra degrees of freedom in
them, and this will have a reflection in the canonical formalism.

\subsection{The Palatini action}
We now write the Einstein action in terms of tetrads. We introduce a
covariant derivative via $D_{a} K_{I} = \partial_{a} K_{I} +
\omega_{aI}{}^{J} K_{J}$. $\omega_{aI}{}^{J}$ is a Lorentz connection (the
derivative annihilates the Minkowski metric). We define a curvature by
$\Omega_{ab}{}^{IJ} = \partial_{[a} \omega_{b]}{}^{IJ} +
[\omega_{a},\omega_{b}]^{IJ}$, where $[\, , \, ]$ is the commutator in the
Lorentz Lie algebra. The Ricci scalar of this curvature  can be expressed
as $e^{a}_{I} e^{b}_{J} \Omega_{ab}^{IJ}$ (indices $I,J$ are raised
and lowered with the Minkowski metric). The Einstein action can be
written,
\begin{equation}
S(e,\omega) = \int d^{4}x\ e\ e^{a}_{I} e^{b}_{J} \Omega_{ab}^{IJ}
\end{equation}
where e is the determinant of the tetrad (equal to $\sqrt{-g})$.

We will now derive the Einstein Equations by varying this action with
respect to $e$ and $\omega$ as independent quantities. To take the
metric and connection as independent variables in the action principle
was first considered by Palatini \cite{Pa}.

As a shortcut to performing the calculation (this derivation is taken
from \cite{Asws}), we introduce a (torsion-free) connection
compatible with the tetrad via $\nabla_{a} e^{b}_{I}=0$. The
difference between the two connections we have introduced is a
field $C_{aI}{}^{J}$ defined by $C_{aI}{}^{J} V_{J}= (D_{a}-\nabla_{a})
V_{I}$. We can compute the difference between the curvatures
($R_{ab}^{IJ}$ is the curvature of $\nabla_{a}$), $\Omega_{ab}{}^{IJ} -
R_{ab}{}^{IJ} =\nabla_{[a} C_{b]}{}^{IJ} + C_{[a}{}^{IM} C_{b]M}{}^{J}$. The
reason for performing this intermediate calculation is that it is
easier to compute the variation by reexpressing the action in terms of
$\nabla$ and $C_{a}{}^{IJ}$ and then noting that the variation with
respect to $\omega_{a}{}^{IJ}$ is the same as the variation with respect
to $C_{a}^{IJ}$. The action, therefore is,
\begin{equation}
S = \int d^{4}x\ e\ e_{I}^{a} e_{J}^{b} (R_{ab}{}^{IJ} + \nabla_{[a}
C_{b]}{}^{IJ} + C_{[a}{}^{IM} C_{b]M}{}^{J}).
\end{equation}
The variation with respect to $C_{a}{}^{IJ}$ is easy to compute: the
first term simply does not contain $C_{a}{}^{IJ}$ so it does not
contribute. The second term is a total divergence (notice that
$\nabla$ is defined so that it annihilates the tetrad), the last term
yields $e_{M}^{[a} e_{N}^{b]} \delta^{M}_{[I} \delta^{K}_{J]}
C_{bK}{}^{N}$. It is easy to check that the prefactor in this expression
is nondegenerate and therefore the vanishing of this expression is
equivalent to the vanishing of $C_{bK}{}^{N}$. So this equation
basically tells us that $\nabla$ coincides with $D$ when acting on
objects with only internal indices. Thus the connection $D$ is
completely determined by the tetrad and $\Omega$ coincides with $R$ (some
authors refer to this fact as the vanishing of the torsion of the connection).
We now compute the second
equation, straightforwardly varying with respect to the tetrad. We
get, (after substituting $\Omega_{ab}{}^{IJ}$ by $R_{ab}{}^{IJ}$ as given
by the previous equation of motion),
\begin{equation}
e_{I}^{c} R_{cb}{}^{IJ} - {\textstyle {1 \over 2}} R_{cd}{}^{MN} e_{M}^{c}
e_{N}^{d} e_{b}^{J}
\end{equation}
which, after multiplication by $e_{Ja}$ just tells us that the
Einstein tensor $R_{ab}-{\textstyle {1 \over 2}} R g_{ab}$ of the metric
defined by the tetrads vanishes.  We have therefore proved that the
Palatini variation of the action in tetradic form yields the usual
Einstein Equations.

There is a slight difference between the first order (Palatini)
tetradic form of the theory and the usual one. One easily sees that a
solution to the Einstein Equations we presented above is simply
$e_{b}^{J}=0$. This solution would correspond to a vanishing metric
and is therefore forbidden in the traditional formulation since
quantities as the Ricci or Riemann tensor are not defined for a
vanishing metric. However, the first order action and equation of
motion are well defined for vanishing triads. We therefore see that
strictly speaking the first order tetradic formulation is a
``generalization'' of General Relativity that contains the traditional
theory in the case of nondegenerate triads. We will see this subtlety
playing a role in the future chapters. It should be noticed that the
possibility of allowing vanishing metrics in General Relativity is
quite attractive since one could evisage the formalism ``going
through'', say, the formation of singularities. It also allows for
topology change \cite{Ho}.

\subsection{The self-dual action}

Up to now the treatment has been totally traditional. We will now take
a conceptual step that allows the introduction of the Ashtekar
variables. We will reconstruct the tetradic formalism of the previous
subsection but we will introduce a change. Instead of
considering the connection $\omega_{a}{}^{IJ}$ we will consider its self dual
part with respect to the internal indices and we will call it
$A_{a}{}^{IJ}$, that is $i A_{a}{}^{IJ}= {\textstyle {1 \over 2}}
\epsilon_{MN}{}^{IJ} A_{a}{}^{MN}$. Now, to really be able to do this, the
connection must be complex (or one should work in an Euclidean
signature). Therefore for the moment being we will consider {\em
complex General Relativity} and we will then specify appropriately how
to recover the traditional real theory. The connection now takes
values in the (complex) self-dual subalgebra of the Lie algebra of the
Lorentz group. We will propose as action,
\begin{equation}
S(e,A) = \int d^{4} x\ e\ e^{a}_{J} e^{b}_{K} F_{ab}{}^{JK}
\end{equation}
where $F_{ab}{}^{JK}$ is the curvature of the self dual connection and
it can be checked that it corresponds to the self-dual part of the
curvature of the usual connection.

We can now repeat the calculations of the previous subsection for the
selfdual case. When one varies the self-dual action with respect to
the connection $A_{a}{}^{IJ}$ again one obtains that this connection
should annihilate the triad (if one repeated step by step the previous
subsection argument, one now finds that the self-dual part of
$C_{a}{}^{IJ}$ vanishes). The variation with respect to the tetrad goes
along very similar lines only that $\Omega_{ab}{}^{IJ}$ gets everywhere
replaced by $F_{ab}{}^{IJ}$. The final equation one arrives to
(exercise) again tells us that the Ricci tensor vanishes.
Remarkably, the self-dual action leads to the (complex) Einstein
Equations. This essentially be understood in the fact that the two
actions differ by boundary terms. We postpone the issue of how to
recover the real theory to subsection 3.6.

\subsection{The New Variables}

If one took the Palatini action of subsection 3.2 and made a
canonical 3+1 decomposition, the formalism basically returns to the
traditional one \cite{AsBaJo}.
A quite different thing happens if one decomposes the
self-dual action. Let us therefore proceed to do the 3+1 split. As
before, we introduce a vector $t^{a}=N n^{a}+N^{a}$ (which, since we
are actually dealing with complex Relativity, may be complex). Taking
the action,

\begin{equation}
S(e,A) = \int d^{4}x\ e\ e_{I}^{a} e_{J}^{b} F_{ab}{}^{IJ}
\end{equation}
and defining the vector fields orthogonal to $n^{a}$, $E^{a}_{I}=
q^{a}_{b} e^{b}_{I}$ (where $q^{a}_{b}=\delta^{a}_{b} + n^{a} n_{b}$
is the projector on the three-surface) we have,
\begin{equation}
S(e,A) = \int d^{4}x\ (e\ E_{I}^{a} E_{J}^{b} F_{ab}{}^{IJ}-2\ e\ E^{a}_{I}
e^{d}_{J} n_{d} n^{b} F_{ab}{}^{IJ}).
\end{equation}
We now define $\tilde{E}^{a}_{I}= \sqrt{q} E^{a}_{I}$, which is a
density on the three manifold. The determinant of the triad can be written
as $e = N \sqrt{q}$. We also introduce the vector in the ``internal
space'' induced by $n^{a}$, defined by $n_{I}=e^{d}_{I} n_{d}$. With
these definitions, and exploiting the self-duality of $F_{ab}{}^{IJ}$ to
write $F_{ab}{}^{IJ} = -i {\textstyle {1 \over 2}} \epsilon^{IJ}{}_{MN}
F_{ab}{}^{MN}$, we get,
\begin{equation}
S(e,A) = \int d^{4}x\ (-{\textstyle {i\over 2}}
\ut{N} \tilde{E}_{I}^{a} \tilde{E}_{J}^{b}
 \epsilon^{IJ}{}_{MN} F_{ab}{}^{MN}-2 N n^{b} \tilde{E}^{a}_{I}
n_{J}  F_{ab}{}^{IJ}).
\end{equation}
We now pick a gauge in which $E^{a}_{0}=0$ and
$n_{I}=(1,0,0,0)$\footnote{One can proceed in a more elegant, albeit
slightly more complicated, way, see ref. \cite{Asws}}. We also exploit
the relation between Levi-Civita densities in three and four space,
$\epsilon^{IJKL} n_{L}=\epsilon^{IJK}$, which in our gauge reads
$\epsilon^{IJK0}=\epsilon^{IJK}$. Therefore, we have,
\begin{equation}
S(e,A) = \int d^{4}x\ (-{\textstyle {i
\over 2}}\ut{N} \tilde{E}_{I}^{a} \tilde{E}_{J}^{b}
 \epsilon^{IJ}{}_{M} F_{ab}{}^{M0}-2 N n^{b} \tilde{E}^{a}_{I}
F_{ab}{}^{I0}).
\end{equation}
Notice that the indices $I,J\ldots{}$ in the above expression now only
range from 1 to 3.
We denote them with lowercase letters to highlight this fact and
rename, $A_{a}{}^{IO}=A_{a}^{i}$ and similarly for
$F_{ab}{}^{IO}=F_{ab}^{i}$. The effect of the presence of the vector in
internal space $n_{I}$ has been to split the two copies of $SU(2)$
present in the Lorentz group in such a way that our new indices
$i,j,k$ are $SU(2)$ indices. We now replace in the second term $N
n^{b}$ by $t^{b}-N^{b}$ and use the identity (exercise) $t^{a}
F_{ab}^{i}= {\cal L}_{t}A_{b}^{i}-{\cal D}_{b}(t^{a} A_{a})^{i}$,
where ${\cal D}_{b}$ is the derivative defined by the connection
$A_{a}^{i}$, to get,
\begin{equation}
S(e,A) = \int d^{4}x\
(2  \tilde{E}^{a}_{i} {\cal L}_{t} A_{a}^{i}
-2 N^{b} \tilde{E}^{a}_{i} F_{ab}^{i}
 -{\textstyle {i \over 2}}\ut{N} \epsilon^{ij}{}_{k} \tilde{E}_{i}^{a}
\tilde{E}_{j}^{b}  F_{ab}^{k}).
\end{equation}
This action is exactly in the form we want. There is a term of the ``$p
\dot{q}$'' form, ($\tilde{E}^{a}_{i} {\cal L}_{t} A_{a}^{i}$), from which we
can read off that the variable canonically conjugate to $A_a^i$ is
$\tilde{E}^a_i$.
The theory also has constraints, given by,
\begin{eqnarray}
\tilde{\cal G}^{i} &=& ({\cal D}_{a} \tilde{E}^{a})^{i} \\
\tilde{C}_{a} &=& \tilde{E}^{a}_{i} F_{ab}^{i} \\
\tilde{\tilde{C}} &=& \epsilon^{ij}{}_{k} \tilde{E}^{a}_{i}
\tilde{E}^{b}_{j} F_{ab}^{k}
\end{eqnarray}
The last four equations correspond to the usual diffeomorphism and
Hamiltonian constraints of canonical General Relativity. The first
three equations  are  extra constraints that stem from our use of triads as
fundamental variables. These equations, which have exactly the same form
as a Gauss Law of an $SU(2)$ Yang-Mills theory, are the generators of
infinitesimal $SU(2)$ transformations. They tells us that the formalism
is invariant under triad rotations, as it should be.

Notice that a dramatic simplification of the constraint equations has
occurred. In particular the Hamiltonian constraint is a polynomial
function of the canonical variables, of quadratic order in each variable.
Moreover, the canonical variables, and the phase space of the theory
are exactly those of a (complex) $SU(2)$ Yang-Mills theory. The reduced
phase space is actually a subspace of the reduced phase space of the
Yang-Mills theory (the phase space modulo the Gauss Law), since
General Relativity has four more constraints that further reduce its
phase space. This resemblance in the formalism to that of a Yang-Mills
theory will be the starting point of all the results we will introduce
in the last sections of this paper.

\subsection{The constraint algebra}

When one has a theory with constraints, one needs to check that these
are consistent with each other, i.e.\  that by  taking Poisson
brackets among the constraints one does not generate new constraints.
If this were the case, these secondary constraints should also be
enforced. Fortunately, the system we have here is first class, i.e.\
the Poisson bracket of each two constraints is a combination of the
other constraints. Actually, in terms of the New Variables, the
structure of the constraints is simple enough for the reader to be
able to compute the constraint algebra without great effort (this
computation can also be carried along with the traditional variables
and the results are the same). We only summarize the results here. To
express them in a simpler form (and to avoid confusing manipulations
of distributions while performing the computations), it is again
convenient to smooth out the constraints with arbitrary test fields.
We denote,
\begin{eqnarray}
{\cal G}(N_{i}) &=& \int d^{3}x N_{i} ({\cal D}_{a} \tilde{E}^{a})^{i}
\\ C(\vec{N}) &=& \int d^{3}x N^{b} \tilde{E}^{a}_{i} F_{ab}^{i} \\
C(\ut{N}) &=& \int d^{3}x \ut{N} \epsilon^{ij}{}_{k} \tilde{E}^{a}_{i}
\tilde{E}^{b}_{j} F_{ab}^{k}
\end{eqnarray}
and as before the notation is unambiguous. The constraint algebra then
reads,
\begin{eqnarray}
\{{\cal G}(N_{i}) , {\cal G}(N_{j}) \} &=& {\cal G}([N_{i},N_{j}])\\
\{C(\vec{N}),C(\vec{M})\} &=& C({\cal L}_{\vec{M}} \vec{N})\\
\{C(\vec{N}),{\cal G}(N_{i})\} &=&{\cal G}({\cal L}_{\vec{N}} N_{i})\\
\{C(\vec{N}),C(\ut{M})\} &=& C({\cal L}_{\vec{N}} \ut{M}) \\
\{C(N_{i}),C(\ut{N})\} &=&0\\
\{C(\ut{N}),C(\ut{M})\} &=& C(\vec{K}) -{\cal G}(A_{a}^i K^{a}),
\end{eqnarray}
where the vector $\vec{K}$ is defined by $K^{a}= 2 \tilde{E}^{a}_{i}
\tilde{E}^{b}_{i} (\ut{N} \partial_{a} \ut{M} - \ut{M} \partial_{a}
\ut{N})$.
Here we clearly see that the constraints are first class. The reader
should notice, however, that the algebra is not a true Lie algebra,
since one of the structure constants (the one defined by the last
equation), is not a constant but depends on the fields
$\tilde{E}^{a}_{i}$ (through the definition of the vector $\vec{K}$).

\subsection{The evolution equations and the reality conditions}

Up to now we have been concerned only with instantaneous relations among
the fields (the constraints). However, if one wants to evolve the
fields in time, one needs the evolution equations, which are simply
obtained taking the Poisson bracket of the fields with the
Hamiltonian.

These equations give the ``time derivative'' of the triad and the connection.
\begin{eqnarray}
\dot{\tilde{E}}^a_i &=& \{\tilde{E}^a_i, H(\ut{N})\} = -i \sqrt{2}
{\cal D}_b (\ut{N} \tilde{E}^{[b} \tilde{E}^{a]})_i \\
\dot{A}_a^i &=& \{ A_a^i , H(\ut{N}) \} =
-{\textstyle {i \over \sqrt{2}}} [ \ut{N}
\tilde{E}^b, F_{ab} ]^i.
\end{eqnarray}
{}From here one can straightforwardly derive the equations of motion for the
traditional variables, the metric and the extrinsic curvature.

We now turn our attention to the issue of ``reality conditions''. As was
mentioned before, the formalism we are dealing with describes complex
General Relativity. In fact, the action we are using is complex! If we
want to recover the classical theory we must take a ``section'' of the
phase space that corresponds to the dynamics of real
Relativity. This can be done. One gives
data on the initial surface that corresponds to a real spacetime and the
evolution equations will keep these data real through the evolution. Now,
strictly speaking, this procedure is not really canonical, since we are
imposing these conditions by hand at the end. That does not mean it is not
useful\footnote{A nontrivial  example where it can be worked out to the
end is the Bianchi II cosmology \cite{GoTa}}.
In fact, one can eliminate the reality conditions and have a
canonical theory. However, many of the beauties of the new formulation are
lost, in particular the structure of the resulting constraints is basically
that of the traditional formalism.

The issue of the reality conditions acquires a different dimension in
the quantum theory. A point of view that is strongly advocated, and
may turn out to be successful, is the following. Start by considering
the complex theory and apply the steps towards canonical quantization
that we discussed subsection (2.4). After the space of physical states
has been found, when one decides to find an inner product, the reality
conditions are used in order to choose an inner product that
implements them. That is, the reality conditions can be a guideline to
find the appropriate inner product of the theory. One simply requires
that the quantities that have to be real according to the reality
conditions of the classical theory, become self-adjoint operators
under the chosen inner product. This solves two difficulties at once,
since it allows us to recover the real quantum theory and the
appropriate inner product at the same time. This point of view is
strictly speaking a deviation from standard Dirac quantization, and
works successfully for several model problems \cite{Ta}. The success
or failure in Quantum Gravity of this approach is yet to be tested and
is one of the most intriguing and attractive features of the
formalism.  (For a critical viewpoint, see \cite{Ku}).

What are, therefore, the reality conditions? They can be written
in several ways, depending on which variables one chooses to express them
in. One could simply write them in terms of the three-metric,
\begin{eqnarray}
&&\tilde{\tilde{q}}^{ab}= (\tilde{\tilde{q}}^{ab})^{*}\\
&&\dot{\tilde{\tilde{q}}}^{ab}= (\dot{\tilde{\tilde{q}}}^{ab})^*
\end{eqnarray}
Using the expressions we presented above for the time derivatives, we
could easily express these equations in terms of the triad and the connection.

How one writes the reality conditions depends largely on what one
wants to accomplish. If one, for instance, is interested in pursuing
the quantization program outlined above, and using the reality
conditions to fix the inner product, one does not necessarily want
them written in the above form. The reason for this is that if one
wants to write them as conditions of hermiticity under an inner
product, one would want to write them in terms of quantities that are
observables of the theory, so that taking their expectation value
makes sense. Since we do not know any observable for the theory, we
cannot write the reality conditions in this way at present.

Let us finish by making a small digression on the issue of
observables. The quantities that one wants to observe in a system are
quantities that are invariant under the symmetries of the theory. Any
other kind of quantity will be gauge dependent and therefore of no
physical relevance. In the canonical theory it is easy to define which
kind of quantities are observables.  Since the constraints are the
infinitesimal generators of the symmetries of the theory, any quantity
that has vanishing Poisson brackets with the constraints is invariant
under the symmetries of the theory. Therefore, if one wanted to find
an observable for General Relativity, one has to look for a quantity
with vanishing Poisson brackets with the diffeomorphism and
Hamiltonian constraint of the theory. The trouble is that we do not
know any single such quantity. This can be a circumstantial problem,
merely reflecting our ignorance, or it could be fundamental. There are
suggestions that maybe no such quantity exists for General Relativity
\cite{AnTo} \footnote{All these remarks refer to the cases of compact
spacetimes. For the asymptotically flat case observables as the four
momentum and angular momentum are well known. Some non-analytic
observables for the compact case may also be written \cite{GoLeSt}}.
This could be related to the fact that the theory displays chaotic
behaviour \cite{FrNe,Pu}.  The issue of observables and their relation
to quantization (some people argue that since the theory may have no
observables, the whole program of quantization we are pursuing here is
doomed) exceeds the scope and is not in line with the emphasis of
these talks so we will not discuss it here. We just want to make the
reader aware that there is potential for a problem with this issue.
For a recent discussion see \cite{An}.  From a practical point of view
one could argue that even if the full theory has no exact observable,
one could find some approximation in which the theory has observables
(after all, we live in such an approximation and measure things all
the time!). Again, this exceeds the scope of this treatment. See
\cite{AsRoSmwe} for more details on this point of view.

\section{Quantum Theory: The Connection Representation}
\subsection{Formulation}
Let us suppose we now decide to proceed and apply the
canonical quantization program to the theory as we have it up to now.

We start by picking a polarization. One has many choices. However, let
us remember that our canonical variables are basically similar to
those of a Yang-Mills $SU(2)$ theory. When one quantizes Yang-Mills
(and Maxwell) theories a usual choice for the polarization is to pick
wavefunctionals of the connection $\Psi(A)$. We will pursue in this
subsection this treatment for General Relativity. Notice that this is
potentially $very$ different from what one does with the traditional
variables, where the more commonly considered polarization is that in
where one takes wavefunctionals of the three metric $\Psi(q)$. In
terms of our variables, this polarization would be closer to choosing
wavefunctionals of the triad. We see that the use of these new
variables leads us to a new perspective even at this level.

A representation for the Poisson algebra of the canonical variables
considered can be simply achieved by representing the connection as a
multiplicative operator and the triad as a functional derivative,
\begin{eqnarray}
\hat{A}_a^i \Psi(A) &=& A_a^i \Psi(A), \\
\hat{\tilde{E}^a_i} \Psi(A) &=& {\delta \over \delta A_a^i} \Psi(A).
\end{eqnarray}

If we now want to promote the constraint equations to operatorial
equations, we need first to pick a factor ordering. Two factor
orderings have been explored, one with the triads to the right (we
will call it II) and one with the triads to the left (I) \footnote{The
reason for this reversal of notation is to keep it in line with the
literature \cite{BrGaPunpb}.} (see \cite{Soo} for alternatives).  Let
us stress that all these calculations are only formal until a
regularization is introduced.

\subsection{Factor ordering II and the role of Wilson Loops}

If one orders the triads to the right, the constraints become,

\begin{eqnarray}
\hat{\tilde{\cal G}}^i &=& D_a {\delta \over \delta A_a^i}\\
\hat{\tilde{\cal C}}_a &=& F_{ab}^i {\delta \over \delta A_b^i}\\
\hat{\tilde{\tilde{\cal H}}} &=& \epsilon^{ijk} F_{ab}^i
{\delta \over \delta A_a^j} {\delta \over \delta A_b^k}
\end{eqnarray}

An attractive feature of this ordering is that the Gauss Law becomes
the infinitesimal generator of $SU(2)$ gauge transformations for the
wavefunctions and the diffeomorphism constraint becomes the
infinitesimal generator of diffeomorphisms on the wavefunctions. This,
among other features, attracted the attention of Jacobson and Smolin
\cite{JaSm} to this ordering. There is a potential awkwardness when
one considers the algebra of constraints. Remember that it is not a
true algebra, but as we discussed, the commutator of two Hamiltonians
has a structure ``constant'' that depends on one of the canonical
variables, the triad. This means that in this ordering such
``constant'' would have to appear to the right of the resulting
commutator, which is not expected, and when one constructs solutions
one has to check the consistency of the constraints \cite{JaSm}.

Jacobson and Smolin set out to find solutions to the constraint
equations in this formalism. If one starts by considering the Gauss
Law, one would like the wavefunctionals to be invariant under $SU(2)$
gauge transformations.  A well known infinite parameter family of
gauge invariant functionals of a connection are the Wilson Loops,
\begin{equation}
W(A,\gamma) = {\rm Tr} \biggr(
{\rm P exp} \oint ds \dot{\gamma}^a(s) A_a(\gamma(s))\biggr),
\end{equation}
defined by the trace of the path-ordered exponential of the line
integral along a loop $\gamma$ (parametrized by $s$) of the
connection. In fact, up to some extent {\em any} Gauge invariant
function of a connection can be expressed as a combination of Wilson
Loops \cite{Barrett,Gi}.  In view of this, one can consider Wilson
loops as an infinite family of wavefunctions in the connection
representation parametrized by a loop $\Psi_\gamma(A) = W(\gamma,A)$
that forms an (overcomplete) basis of solutions to the quantum Gauss
Law constraint. So, we have managed to find solutions to the first set
of constraints, even perhaps a basis of solutions.

What happens to the diffeomorphism constraint? Evidently Wilson loops
are not solutions. When a diffeomorphism acts on a Wilson loop, it
gives as a result a Wilson loop with the loop displaced by the
diffeomorphism performed.  Therefore they are not annihilated by the
diffeomorphism constraint and cannot become candidates for physical
states of Quantum Gravity.  In spite of that, they are worthwhile
exploring a bit more. Remember they form an overcomplete basis in
terms of which any physical state should be expandable (since any
physical state has to be Gauge invariant). We will therefore explore
what happens when we act with the Hamiltonian constraint on them. To
perform this calculation we only need the formula for the action of a
triad on a Wilson Loop,
\begin{equation}
\hat{\tilde{E}}^a_i(x) \Psi_\gamma(A) = {\delta \over \delta A_a^i(x)}
\Psi_\gamma(A)
=\oint ds \ \delta^3(x-\gamma(s)) \dot{\gamma}^a(s) {\rm Tr}(U(0,s) \tau^i
U(s,1)) \label{fundamental}
\end{equation}
where we denote by $U(s_1,s_2)= \int_{s_1}^{s_2}
dt \dot{\gamma}^a(t) A_a(\gamma(t))$
the
holonomy from the point $\gamma(s_1)$ to the point $\gamma(s_2)$;
$\tau^i$ denotes
a Pauli matrix. Note that
this expression involves a line integral of a three-dimensional delta
function. Using some notational latitude, we can rewrite it as,
\begin{equation}\label{functdonw}
\hat{\tilde{E}}^a_i(x) \Psi_\gamma(A) = \dot{\gamma}^a(x) {\rm Tr}(U(0,s(x))
\tau^i U(s(x),1)).
\end{equation}
The notational latitude consists in the fact that we have
``cancelled'' a three dimensional Dirac Delta with a one dimensional
integral (which hides a regularization problem) and we have denoted by
$s(x)$ the parameter value $s$ for which the loop is at the point $x$
(one has to be careful with this notation if the loop multiply
traverses such a point, as in the case of an intersection). The point
for this notational deviation is to make more transparent the
following result (which actually goes through even regularizing with
some care \cite{JaSm}, up to the extent that that is possible in this
context!). Let us evaluate the action of the Hamiltonian constraint on
a Wilson Loop. Using the previous formulae we get,
\begin{eqnarray}
\hat{H}(x) \Psi(A) &=& \epsilon^{ijk}F_{ab}^i(x) {\delta \over \delta A_a^j(x)}
{\delta \over \delta A_b^k(x)} \Psi(A) = \\
&=& F_{ab}^i
\dot{\gamma}^a(x) \dot{\gamma}^b(x) {\rm Tr}(U(0,s(x))\tau^i U(s(x),1))
\nonumber
\end{eqnarray}
Notice that in this expression we have an antisymmetric tensor,
$F_{ab}^i(x)$, contracted with a symmetric tensor $\dot{\gamma}^a(x)
\dot{\gamma}^b(x)$.  Therefore, the expression vanishes! We have just
proved that a Wilson loop formed with the Ashtekar connection is a
solution of the Hamiltonian constraint of Quantum Gravity. This is a
remarkable fact. Notice that up to this discover {\em no} solution of
this constraint was known in a general case (without making
minisuperspace approximations). Historically, this discovery fostered
the interest for loops in this context and led to the use of the loop
representation.

There are some drawbacks to this result. An obvious one is that
although we solved the Hamiltonian constraint and the Gauss Law, we
did not solve the diffeomorphism constraint, therefore these
wavefunctions are not states of Quantum Gravity. The second point is
that for the above result to hold, we need the tensor
$\dot{\gamma}^a(x) \dot{\gamma}^b(x) $ to be symmetric. This is true
if we have a smooth loop. If the loop has kinks or intersections, this
is not any longer true and the Wilson loops stop to be solutions (at
this naive level).  Why care about loops with intersections? Why not
just restrict ourselves to smooth loops?  The problem appears when we
try to get some sort of understanding of what these wavefunctionals
are. The first question that comes to mind (of a prejudiced relativist
at least) is what is the metric for such a state. This in principle is
a meaningless question, since the metric is not an observable, but let
us ask it anyway to see where it leads.  The metric acting on one of
these states, gives,
\begin{equation}
\hat{\tilde{\tilde{q}}}^{ab}(x) \Psi_\gamma(A) =
{\delta \over \delta A_a^i}{\delta \over \delta A_b^i} \Psi_\gamma(A) =
\dot{\gamma}^a(x) \dot{\gamma}^b(x) \Psi_\gamma(A)
\end{equation}
So Wilson Loops constructed with smooth loops are eigenstates of the
metric operator.  First, notice that the metric only has support
distributionally along the loop (our sloppy notation may not make this
totally transparent, but notice that the quantity $\dot{\gamma}^a(x)$
only can be nonvanishing along the loop).  Then, notice that the
metric has only one nonvanishing component, the one along the loop.
Therefore it is a degenerate metric. Now, this statement is still
meaningless in a diffeomorphism invariant context, but it actually can
be given a rigorous meaning with a little elaboration.  Consider the
quantum operator obtained by computing the (square root of the)
determinant of the three metric.  In terms of the new variables it is
given by $\epsilon^{ijk} \ut{\epsilon}_{abc} \tilde{E}^a_i
\tilde{E}^b_j \tilde{E}^c_k$. It is very easy to see, as is expected
for a degenerate metric, that this operator vanishes when applied to a
Wilson Loop. The problem with this is made clear when we consider
General Relativity with a cosmological constant. The only thing that
changes in the canonical formalism is that the Hamiltonian constraint
gains an extra term,
\begin{equation}
H_\Lambda = H_0 + \Lambda {\rm det} q
\end{equation}
where $H_0$ is the vacuum Hamiltonian constraint and $\Lambda$ is the
cosmological constant. The extra term is given by the determinant of
the three metric. Now consider our Wilson Loop state. Since it is
annihilated by the vacuum Hamiltonian constraint {\em and} the
determinant of the three metric, this means it is a state for an
arbitrary value of the cosmological constant! That spells serious
trouble. General Relativity with and without a cosmological constant
are totally different theories, and one does not expect them to share
a common set of states, except for special situations, as for
degenerate metrics.

It turns out one can improve the situation a little using
intersections. One can find some solutions to the Hamiltonian
constraint even for the intersecting case by combining holonomies in
such a way that the contributions at the intersection cancel
\cite{JaSm,Hu,BrPu91}.  However, unexpectedly, this is not enough to
construct nondegenerate solutions. All the solutions constructed in
this fashion, if they satisfy the Hamiltonian constraint, are also
annihilated by the determinant of the metric \cite{BrPu91}.  This,
plus the fact that they do not satisfy the diffeomorphism constraint,
shows that these solutions are of little physical use in this context.
They were, however, very important historically as motivational
objects for the study of loops. We will show later on how, when one
works in the loop representation, it is possible to generate solutions
to all the constraints that, although still based on loops, do not
have this degeneracy problem.

\subsection{Factor ordering I: The role of the Chern-Simons form}

If one orders the constraints with the triads to the left, there is
potential for a problem: as we said, apparently in this factor
ordering the diffeomorphism constraint fails to generate
diffeomorphisms on the wavefunctions. For many of us, this would be a
reason to abandon this ordering altogether. However, by considering a
very generic regularized calculation one can prove that the diffeo
constraint actually generates diffeomorphisms, so this is not a
problem \cite{BrGaPunpb}.  Besides, there is the advantage that when
one considers the constraint algebra, one obtains (these are only
formal unregulated results) the correct closure \cite{As}.

In this ordering, Wilson Loops do not solve the Hamiltonian constraint
anymore.  However, there is a very interesting and rich solution one
can construct.  Consider the following state, function of the
Chern-Simons form built with the Ashtekar connection,
\begin{equation}
\Psi_{\Lambda}[A] = {\rm exp} (-{\textstyle{6 \over \Lambda}}
\int \tilde{\epsilon}^{abc} Tr[A_a
\partial_b A_c +{\textstyle {2 \over 3}} A_a A_b A_c])
\end{equation}
This functional has the property that the triad (you can view it as an
electric field) equals the magnetic field formed from the Ashtekar
connection.
\begin{equation}
{\delta \over \delta A_a^i} \Psi_{\Lambda}[A] = {\textstyle {6 \over \Lambda} }
\tilde{\eta}^{abc} F_{bc}^i \Psi_{\Lambda}[A]\label{functdoncs}
\end{equation}
Besides, it is well known that this functional is invariant under
(small) gauge transformations and diffeomorphisms. One can check that
it is annihilated by the corresponding constraints. What may come as a
surprise is that it is actually annihilated by the Hamiltonian
constraint with a cosmological constant. This is easy to see, simply
consider the constraint,
\begin{equation}
\hat{\cal H} = \epsilon_{ijk} {\delta \over \delta A_a^i}
{\delta \over \delta A_b^j} F_{ab}^k -{\Lambda \over 6}
\epsilon_{ijk} \ut{\epsilon}^{abc}
{\delta \over \delta A_a^i}
{\delta \over \delta A_b^j}
{\delta \over \delta A_c^k}
\end{equation}
and notice that the rightmost derivative of the determinant of the
metric simply reproduces the term on the left when acting on the
wavefunction.  Notice that the result holds without even considering
the action of the other derivatives, and therefore is very robust vis
a vis regularization.  This result was independently noticed by
Ashtekar \cite{Aspoo} and Kodama \cite{Ko}. A nice feature of this
result is that the metric is nondegenerate in the sense that we
discussed in the previous section. The metric is just given by the
trace of the product of two magnetic fields. Such property holds
classically for spaces of constant curvature. This has lead some
authors to suggest this wavefunction as a ``ground state'' for a
DeSitter geometry \cite{Korev}.

Now, this does not seem a very impressive feat. First of all, it is
only one state. Secondly a similar state is present in Yang-Mills
theory (this is easy to see, since the Hamiltonian is $E^2+B^2$ and
adjusting constants one gets for the corresponding state $E=iB$) and
is known to be nonphysical since it is nonnormalizable.  This is true,
but it is also true that the nature of a theory defined on a fixed
background as Yang Mills theory is expected to be radically different
from that of a theory invariant under diffeomorphisms, as General
Relativity.  Therefore normalizability under the inner product of one
theory does not necessarily imply or rule out normalizability under
the inner product of the other.  It is remarkable that the
Chern-Simons form, which is playing such a prominent role in particle
physics nowadays, should have such a singular role in General
Relativity. It is the only state in the connection representation that
we know that may have something to do with a nondegenerate geometry!!

There are more things one could say about the connection
representation. There is the compelling work of Ashtekar, Balachandran
and Jo \cite{AsBaJo} concerning the CP violation problem and the
partial success (in the linearized theory) of Ashtekar \cite{Asog} in
addressing the issue of time.  We do not have space here to make
justice to these pieces of work and we refer the reader to the
relevant literature.

\section{The Loop Representation}
\subsection{Motivation}

A rigorous formulation of a loop representation for a quantum field
theory is elaborate and involves many delicate details.  However, one
can get a very simple intuitive grasp of the idea if one ignores
subtleties, and that is exactly what we will do here. We will mention
some of the problems, but the reader who wants to get an idea of the
subtleties is encouraged to see references \cite{AsIs,Le}.

The way to easily view the concept of a Loop Representation is to
compare it to the Momentum Representation of ordinary Quantum
Mechanics.  In the latter, one starts from a wavefunction in the
Position Representation $\Psi(\vec{x})$, and convolutes it with an
(infinite) basis of functions parametrized by a continuous parameter
$\vec{k}$, ${\rm exp}(i \vec{k}\cdot\vec{x})$. The result depends on
the continuous parameter $\vec{k}$ and we call it a wavefunction in
the momentum representation,
\begin{equation}
\Psi(\vec{k}) =  \int d^3x \  {\rm exp}(i \vec{k}\cdot\vec{x}) \Psi(\vec{x}).
\end{equation}
Now, in the Loop Representation, we start from a wavefunctional in the
connection representation $\Psi(A)$. We convolute it with an infinite
basis of (gauge invariant) wavefunctions parametrized by a continuous
parameter, a loop $\gamma$, the Wilson Loops $W_\gamma(A)$. The result
depends on the parameter $\gamma$ and is a wavefunction in the Loop
Representation,
\begin{equation}\label{transform}
\Psi(\gamma) = \int ``d A'' W(A,\gamma) \Psi(A).
\end{equation}
The reader may have several reservations at this point. First of all,
what is the measure to perform this integral? We do not know and we
denoted that by the quotation marks. Secondly, up to what extent is
this transform complete or how faithfully can it represent the
wavefunction space. Again we do not quite know the answer. These
technical mathematical issues have been addressed up to some extent in
the literature and we will not discuss them here. For us to work, it
suffices to notice that this transform works for several model
problems, as 2+1 gravity \cite{AsHuRoSaSm}, Maxwell theory \cite{AsRo}
and Chern-Simons theory \cite{Li}.  Better yet, it works for
Yang-Mills on a lattice \cite{GaLeTr,Ar,Br}, which includes the phase
space of Gravity in terms of the New Variables. For the case of real
nonabelian connections, the transform can be given rigorous meaning
\cite{AsIs}.

Historically, the first construction of a loop representation for
Quantum Gravity based on Ashtekar's new variables is due to Rovelli
and Smolin \cite{RoSmprl,RoSmnpb}, immediately following the results
with Wilson Loops in the connection representation by Jacobson and
Smolin \cite{JaSm}.  In the context of gauge theories, and even
gravity in terms of traditional variables, loop representation had
been considered before by Gambini and Trias \cite{GaTr}.

Another point that may disquiet the reader is how can one know or at
least guess that the quantity of information conveyed in a connection
on a three manifold is equal to that present in the set of all
possible loops on the manifold. Leaving aside technicalities, it turns
out that the loop basis is way overcomplete. The practical
inconvenience of this fact can be clearly seen, for example, in
investigations done with the loop representation on the lattice.
Formulating exactly what are the ``free'' degrees of freedom in the
loop representation is an open and difficult problem \cite{Lo}.  There
are lots of hidden identities among states in the loop representation.
They can all be derived from three basic identities among Wilson
loops:

\begin{itemize}
\item It is easy to see that reversing the
orientation of the loop leaves the Wilson loop invariant. Therefore
wavefunctions in the Loop Representation are invariant under reversal
of orientation of the loop $\Psi(\gamma)=\Psi(\gamma^{-1})$.
\item Due to the fact that the Wilson Loop is the trace of a holonomy and
that the traces are cyclic, $\Psi(\gamma_1 \circ \gamma_2) = \Psi(\gamma_2
\circ \gamma_1)$.
\item There is a nontrivial identity satisfied between traces of $SU(2)$
matrices (other groups have similar identities, although there are
different from the specific one for $SU(2)$), ${\rm Tr}(A){\rm
Tr}(B)={\rm Tr} ( A \cdot B)+ {\rm Tr}(A \cdot B^{-1})$, where $A,B$
are $SU(2)$ matrices.  This is usually called the Mandelstam identity.
In terms of the Loop Representation, this means we can express any
wavefunctional of two loops $\Psi(\gamma_1,\gamma_2)$ in terms of
wavefunctionals of one loop by $\Psi(\gamma_1,\gamma_2)=\Psi(\gamma_1
\circ \gamma_2) + \Psi(\gamma_1 \circ \gamma_2^{-1})$. From now on
therefore, we need only concentrate on wavefunctionals of a single
loop, the multiple loop cases always being reducible to the single
loop instance\footnote{The Mandelstam identity obviously holds only
for a pair of intersecting loops $\gamma_1, \gamma_2$. We are
considering loops that share a common basepoint, that is why we can
always reduce a wavefunction to that of one loop}.
\end{itemize}

Unfortunately, the story does not end here. Many nontrivial
combinations of these identities are possible, for example the
following identity holds for three loops,
\begin{equation}
\Psi(\gamma_1 \circ \gamma_2 \circ \gamma_3) +
\Psi(\gamma_1 \circ \gamma_2 \circ \gamma_3^{-1}) =
\Psi(\gamma_2 \circ \gamma_1 \circ \gamma_3)+
\Psi(\gamma_2 \circ \gamma_1 \circ \gamma_3^{-1})
\end{equation}
(and usually it takes some time to figure out exactly how to derive it
from the above identities). The situation only gets worse for higher
numbers of loops. An important point, however, is that any
wavefunction on loop space that one wants to define (any wavefunction
must have some value for a given number of loops) should be {\rm
consistent} with these identities.

There is a very important aspect of the Loop Representation for
Quantum Gravity. Since the theory is invariant under diffeomorphisms,
this means that the wavefunctionals will be invariant under
deformations of the loops.  This means the wavefunctionals will be
what in the mathematical literature is known as {\em knot invariants}
\cite{RoSmprl}.  Knot theory is the branch of mathematics that studies
properties of knot invariants and it has enjoyed a great deal of
activity recently. It is really exciting that this new flourishing
branch of mathematics seems to have something to do with Quantum
Gravity, since it opens the opportunity for new insights into the
field. We will return to these issues at length in the next chapter.

\subsection{A more rigorous approach}

We said that the introduction of a loop representation via a transform
is only a motivational approach since one actually does not know how
to perform the functional integral appearing in the transform. It
turns out there is a more rigorous way of introducing a loop
representation, and this is by quantizing a noncanonical algebra of
loop-dependent quantities.  This approach was followed for gravity by
Rovelli and Smolin in their original article \cite{RoSmprl}. We will
not discuss it in great detail here for reasons of space. The main
idea is the following:
\begin{itemize}
\item Define a set of classical quantities in the phase space of General
Relativity based on loops, called generically T variables. $T^0_\gamma
[A]$ is the Wilson Loop, $T^1$ is a vector density obtained by
inserting in the expression of the Wilson loop a triad at a given
point $T^{a}(x)_\gamma [A] = Tr(U(0,s(x))\tilde{E}^a U(s(x),1))$,
$T^2$ is a two-vector density obtained by inserting in a Wilson loop
triads at two points and so on for higher order T's. It is possible,
by shrinking the loops to points, to represent any classical quantity
in terms of these variables. In particular, one can write the
constraints. These quantities close a noncanonical algebra under
Poisson Brackets.
\item Promote the T variables to a set of operators acting on a space of
wavefunctions. The action is obtained by mimicking the action under
Poisson brackets of the classical T variables with the Wilson loop.
For example, $\hat{T^0_\gamma} \Psi(\eta)= \Psi(\gamma \circ \eta)$
and so on.  The quantum operators close an algebra under commutators
that reproduces, in the limit $\hbar \to 0$ the classical T algebra.
\item Promote any quantity one is interested in (for example the constraints)
to quantum operators simply by writing its classical expression in terms
of the T variables and then promoting them to operators by the rules given
above.
\end{itemize}

There are subtleties in the definition of the quantum algebra of T
variables that have led to the consideration of strips instead of loops
to avoid some regularization issues. See Rovelli \cite{Ro} for details.

\subsection{Differential operators in loop space} If one accepts that
a representation of Quantum Gravity exists in which wavefunctions are
functionals of loops, the next question is how to represent the
constraint equations. To this end we will introduce a differential
operator in Loop Space, the loop derivative. Given a functional of a
loop $\Psi(\gamma)$, the loop derivative is defined by,
\begin{equation}\label{loopd}
\Psi(\gamma\circ\delta\gamma) = (1 +\sigma^{ab} \Delta_{ab}(P) ) \Psi(\gamma)
\end{equation}
\begin{figure}[b]
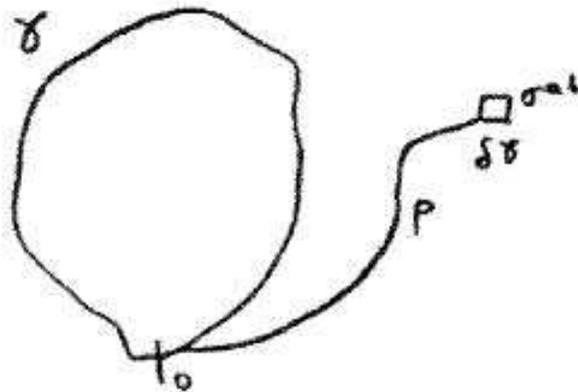

\vskip 5cm
\caption{The infinitesimal loop that gives rise to the Loop Derivative}
\label{figaread}
\end{figure}
where $\delta \gamma$ is an infinitesimal loop added at the end of the path
P as indicated in the figure \ref{figaread}.
$\sigma^{ab}$ is the area  of the
infinitesimal loop.  The prescription to compute it is, take the wavefunction
evaluated on a loop constructing by appending to the original loop an
infinitesimal loop $\delta \gamma$ at the end of a path P starting from the
basepoint; subtract the wavefunction evaluated on the original loop, and
divide the result by the area element $\sigma^{ab}$. The result is the
loop derivative of the wavefunction. The loop derivative is present in the
work of Mandelstam \cite{Ma}, Polyakov \cite{Po} and Makeenko and Migdal
\cite{MaMi}. The work of Gambini and Trias really brought to the forefront
the role of this operator in connection with gauge theories \cite{GaTr}.

A fundamental property that can be immediately checked applying the above
prescription is,
\begin{eqnarray}
\Delta_{ab}(\gamma_0^x) W(\gamma,A) &=& Tr(U(0,s(x)) F_{ab}(x) U(s(x),1))
\label{loopdf}\\
&=& F_{ab}^i Tr(U(0,s(x)) \tau^i U(s(x),1)).\nonumber
\end{eqnarray}
 As usual, one should be careful if such a point lies at an intersection.
However, in most practical situations, intersections are regulated and
derivatives act away from them and are evaluated at the intersection only
in the limit.
\subsection{The diffeomorphism constraint}
In terms of the loop derivative it is straightforward to write the generator
of infinitesimal diffeomorphisms. Although we could proceed from geometric
considerations, let us derive it from the  New Variable formulation.
Let us consider the generator of diffeomorphisms acting on both sides of
the transform (\ref{transform}),
\begin{equation}
\hat{C}(\vec{N}) \Psi(\gamma) = \int d A\ W(\gamma,A)  \int d^3 x N^a(x)
{\delta \over \delta A_b^i(x)} F_{ab}^i(x) \Psi(A)
\end{equation}
we now integrate by parts and apply the constraint on the Wilson Loop,
\begin{equation}
\hat{C}(\vec{N}) \Psi(\gamma) = \int d A \int d^3 x N^a(x) F_{ab}^i(x)
{\delta \over \delta A_b^i(x)} W(\gamma,A) \Psi(A)
\end{equation}
We now functionally differentiate the Wilson Loop and use the equations
(\ref{fundamental}) and (\ref{loopdf}),
\begin{eqnarray}
&& F_{ab}^i(x) {\delta \over \delta A_b^i(x)} W(\gamma,A) =
F_{ab}^i(x) \dot{\gamma}^b(s(x)) Tr(U(0,s(x)) \tau^i U(s(x),1)=\\
&& = \dot{\gamma}^b(x) Tr(F_{ab}(x) U(x,x)) = \dot{\gamma}^b(x) \Delta_{ab}(x)
W(\gamma,A)
\end{eqnarray}
Therefore we can now replace this in the expression of the constraint to
find,
\begin{equation}
\hat{C}(\vec{N}) = \int d^3 x N^a(x) \oint ds  \delta(x-
\gamma(s)) \dot{\gamma}^b(s) \Delta_{ab}(x)
\end{equation}
Whenever we drop the dependence in the path of the loop derivative, we assume
the path goes from the basepoint of the loop to the point of interest.
The expression we arrived to is known to be the generator of infinitesimal
deformations of the loops.
What we see above is what we anticipated at the end of the last subsection,
that the wavefunctions will have to be invariant under smooth deformations
of the loops, they have to be {\em knot invariants}.

An important point if we are really to consider the above expression as a
diffeomorphism constraint is if it reproduces the constraint algebra of
diffeomorphisms. The calculation is possible, though we will not perform
it here for reasons of space. The interested reader can consult
\cite{GaTr,GaGaPu}.

\subsection{The Hamiltonian constraint}

We can now use the same kind of reasoning to determine the action of
the Hamiltonian constraint. Naturally, it will be more complicated.
Again for reasons of space we will not perform a derivation here.
Careful derivations can be found in \cite{Ga,BrPushr}. The resulting
expression is,
\begin{eqnarray}
&&\hat{H}(x) \Psi(\gamma) = \oint_\gamma ds \oint_\gamma dt \
\delta(x-\gamma(s)) f_\epsilon(\gamma(s),\gamma(t)) \times \\
&&\times
\dot{\gamma}^a(s) \dot{\gamma}^b(t) \biggr( \Delta_{ab}(s) \Psi(\gamma_{st}
\circ \gamma_{s0t}) + \Delta_{ab}(s) \Psi(\gamma_{ts} \circ \gamma_{t0s})
\biggr)\label{hamiltonian}
\end{eqnarray}
This expression requires some explanation. Conceptually it can be viewed as
``$\dot{\gamma}^a \dot{\gamma}^b \Delta_{ab}$''. However, several details
should be taken into account. First of all, the expression is regulated with
a regulator satisfying $f_\epsilon(x,y) \to \delta(x-y)$ when $\epsilon \to
0$. This means the two tangent vectors effectively are evaluated at the same
point. Since they are contracted with the loop derivative, which is
antisymmetric, this means that the only possibility for the constraint to
be nonvanishing is if it acts at an intersection. This fact we already
encountered in the connection representation. The parameter values
$s$ and $t$ can therefore be thought as slightly displaced from an
intersection (which lies at $x$), but tend to it as we remove the
regulator $\epsilon \to 0$. There is a rerouting in the loops, indicated in
the wavefunctions. For instance, this should be read in the following way;
$\Psi(\gamma_{st} \circ \gamma_{t0s})$ should be understood as the loop
composed by traversing the original loop from $s$ to $t$ first and then
from $t$ to $s$, but going through the basepoint $0$. This is depicted
in the figure \ref{reroutings}.
\begin{figure}[h]
\vskip 5cm
\caption{The loops $\gamma$, $\gamma_{st}\circ \gamma_{s0t}$ and
$\gamma_{ts} \circ \gamma_{t0s}$.}
\label{reroutings}
\end{figure}
The reader may find this expression technically difficult. However, it
should be kept in mind that this expression embodies all the dynamical
information of the General Theory of Relativity at a quantum level in
loop space. When viewed in this way the above expression appears as
remarkably simple!

At this point the reader may be puzzled about the kind of derivations
we have performed for the constraints. We have made use of the loop
transform (\ref{transform}) as if it were an actually well defined
expression, instead of the motivational tool we claimed it to be in
the introduction. There actually is a way of deriving the form of the
constraints in the loop representation without making use at all of the
transform (\ref{transform}). One takes the second viewpoint we described
in connection with the construction of the loop representation in section
5.2: its
elaboration based on a quantization of a noncanonical set of loop-dependent
classical quantities (the Rovelli-Smolin T operators). The idea is to
express the constraints classically in terms of the T's. Then, since one
has a well defined way of promoting the T operators to quantum operators on
a space of loop functionals, one has also a prescription for promoting
the constraints to differential operators in loop space. It is quite
reassuring that proceeding in this way and proceeding via the transform
as we detailed above, actually leads to the same results for both the
Hamiltonian and diffeomorphism constraints \cite{BrPushr}.

\subsection{Coordinates on loop space}
We have established analytic expressions for the constraints of Quantum
Gravity in the Loop Representation. We now develop technology that will
allow us to write wavefunctions. We will discuss specifically the
construction of wavefunctions in the next chapter. Here we will just set
up the framework that will be used.

Let us return for a minute to the connection representation. We are interested
in wavefunctions that are $SU(2)$ invariant. Modulo technicalities, all such
functions can be expressed as combinations of Wilson Loops \cite{Ba,Gi}.
Therefore we
can just concentrate on these latter ones. Let us write the expression for
a Wilson Loop explicitly,
\begin{equation}
W(\gamma,A) = 2 +\sum_{i=1}^\infty {\rm Tr}[ \oint ds_1 \int_0^{s_1}
ds_2 ...\int_0^{s_{n-1}} ds_n \dot{\gamma}^{a_1}(s_1)...
\dot{\gamma}^{a_n}(s_n) A_{a_1}(\gamma(s_1))...A_{a_n}(\gamma(s_n))]
\end{equation}
We now rearrange this expression in the following way,
\begin{eqnarray}
&&W(\gamma,A) = 2 + \sum_{i=1}^\infty
\int d^3 x_1 ... \int d^3 x_n {\rm Tr}(A_{a_1}(x_1)...A_{a_n}(x_n)) \times \\
&&\times\oint ds_1...\oint ds_n \Theta(s_1,...,s_n)
\delta^3(x_1 - \gamma(s_1))...
\delta^3(x_n - \gamma(s_n)) \dot{\gamma}^{a_1}(s_1)...\dot{\gamma}^{a_n}(s_n)
\end{eqnarray}
where $\Theta(s_1,...s_n)=1$ if $s_1<...<s_n$ is a generalized Heaviside
Function.
The purpose of this rearrangement is to separate the contributions of the
loop and the connection to the Wilson Loop. This allows us to write,
\begin{equation}\label{Wilson}
W(\gamma,A) = 2 + \sum_{n=1}^\infty
{\rm Tr}(A_{a_1 x_1}...A_{a_n x_n}) X^{a_1 x_1...a_n x_n}(\gamma)
\end{equation}
where,
\begin{eqnarray}
X^{a_1 x_1...a_n x_n}(\gamma) = \oint ds_1...\oint ds_n \
\dot{\gamma}^{a_1}(s_1)...
\dot{\gamma}^{a_n}(s_n) \times \nonumber\\
\times \Theta(s_1,...,s_n) \delta^3(x_1-\gamma(s_1)) ...
\delta^3(x_n-\gamma(s_n))\label{coord}
\end{eqnarray}
where we have assumed a ``generalized Einstein convention'' meaning repeated
$x_i$ coordinates are integrated over and we treat them as indices. We have
isolated all the loop dependence in the quantities $X$. These quantities
behave like multi vector densities (really they are distributions) on the
three manifold at the points $x_i$. They can be viewed as vector densities
that have support along the loops associated with the value of the
tangent vector to the loop at the point of interest.

The whole point of this construction is that these quantities embody
all the information of the loops that is needed to build {\em any}
wavefunction of interest. This transcends the connection
representation and means we can use them in the loop representation.
If we are to write functions of the $X's$ as candidates for physical
states of gravity, we will need to study the action of the constraints
on the $X's$. We do not have space here to give a detailed account of
this, but we will work out an example just to show how this works. The
reader should be able to generalize to the other needed cases.

Let us consider the multitangent of order one, $X^{ay}$, and let us
study its loop derivative. We start by the definition (\ref{loopd}),
appending an infinitesimal parallelogram loop (at the point z) with edges
$du^a$, $dv^a$ to the definition of the $X$ of order one
(\ref{coord}),
\begin{eqnarray}
&&X^{a y}(\gamma \circ \delta \gamma_z)= \int_0^z ds\  \dot{\gamma}^a(s)
\delta(\gamma(s)-x) + du^a \delta(z-x) +dv^a \delta(z+du-x) -\nonumber\\
&&-du^a \delta(z+du+
dv-x) -dv^a \delta(z+dv-x) +\int_z^1 ds\ \dot{\gamma}^a(s)
\delta(\gamma(s)-x)
\end{eqnarray}
where  we now expand to first order assuming $du$ and $dv$ are
infinitesimal (for instance, $\delta(z+du-x)=\delta(z-x)+du^a \partial_a
\delta(z-x)$, and we also recombine the first and last terms to give
a loop integral and get,
\begin{equation}
X^{a x}(\gamma \circ \delta \gamma_z)=X^{a x}(\gamma) +(du^a dv^b -dv^a du^b)
\partial_b \delta(x-z).
\end{equation}
We now read off from the definition of the loop derivative (\ref{loopd}),
taking into account that the area element of the loop in question is given by
$d\sigma^{ab} =du^a dv^b -dv^a du^b$,
\begin{equation}
\Delta_{cb}(z) X^{a x}(\gamma) = \delta^a_{[c} \partial_{b]} \delta(x-z)
\end{equation}
Similar computations can be carried along for the higher order $X's$, for
instance for a second order one,
\begin{equation}
\Delta_{cd}(z) X^{a x b y} = X^{b y} \delta^a_{[c} \partial_{d]} \delta(x-z) +
X^{a x} \delta^b_{[c} \partial_{d]} \delta(y-z) +\delta^a_{[c} \delta^b_{d]}
\delta(x-z) \delta(x-y)
\end{equation}
Generic formulae for an X of any order are given in \cite{DiGaGrLe}.

The reader may be surprised by the fact that we use the word ``coordinates''
to describe the multitangents. A ``coordinate'' should be an object that one
prescribes freely. How do we know we can freely prescribe the X's to any
order? Actually we cannot. The X's satisfy a series of identities, both
algebraic and differential. Examples of these identities are,
\begin{eqnarray}
\partial_a X^{ax} =0\\
\partial_{ax} X^{ax\,by} = \delta(x-y) X^{by}\\
X^{ax\,by} + X^{by\,ax} = X^{ax} X^{by}
\end{eqnarray}
The algebraic identities, like the last one, stem from identities of the
generalized Heaviside Function. The differential identities ensure that
the resulting Wilson loop is gauge invariant. Therefore the X's are not
coordinates, since they cannot be freely specified.

A remarkable fact, however, is that one can actually solve the aforementioned
identities for the ``freely specifiable part'' of the loop coordinates (in
reference \cite{DiGaGrLe}, they are referred to as Y's). These objects
really work as coordinates in loop space. The importance of this last
fact cannot be overstressed. Unfortunately, for reasons of space we are
unable to present a thorough account of this construction here. The
reader is referred to the literature for more details \cite{DiGaGrLe}.
We will keep on to loosely refer to the X's as loop coordinates.

Another important point connected with the loop coordinates is that they
suggest a way to obtain a quantum representation that goes beyond the
loop representation. Consider a generic multivector density on a
three manifold, X, not necessarily associated with any particular loop.
If this quantity satisfies the family of identities mentioned above,
one could use it in expression (\ref{Wilson}) and obtain as a result,
not a Wilson loop anymore, but a gauge invariant function of the connection
parametrized by the vector density given. This would allow, for example,
to use completely smooth vector densities, that have potential for
removing several of the regularization difficulties associated with the
loop representation. We call the resulting representation, where
wavefunctions are functionals of the coordinates,
\begin{equation}
\Psi(X^1, X^2,...),
\end{equation}
``coordinate representation''
(some authors call it ``form factor representation''). At least for the
Maxwell case it has been proven that this provides a viable quantum
representation \cite{AsRo}. For the nonabelian case, work is in progress
to assess the feasibility of this representation.

\section{Knot Theory} \subsection{Knots} Knot theory studies the
properties of knots in three dimensions that are invariant under
smooth deformations of the knots (diffeomorphisms). It was largely set
up by the attempts in the last century by P. G. Tait \cite{Tait} and
others to explain the properties of atoms using knotted vortices of
``aether'' (this was before special Relativity or Quantum Mechanics
were invented. A discussion of several appealing aspects of this
theory of atomic spectra can be found in the book by Atiyah
\cite{At}).  We will use the term ``knot'' loosely to refer either to
a single knotted curve or to several curves linked and/or knotted.  A
central issue in knot theory is to distinguish and classify
inequivalent knots. A useful tool for accomplishing this is the use of
knot invariants, that is, numbers associated with knots which are
invariant under deformations of the knots.  At the moment, however,
there is no ``canonical'' or definite set of invariants that would
allow us to completely classify knots.  The simplest example of such
an invariant is the linking number, first considered by Gauss.
Consider two curves as in figure \ref{hopf}, the linking number is one
if they are linked (\ref{hopf}a) and zero if they are not linked
(\ref{hopf}b).
\begin{figure}[b]
\vskip 5cm
\caption{The knots in  (b) and (c) (The Whitehead
Link) are not linked}
\label{hopf}
\end{figure}
It is evidently (at least for these cases)
invariant under diffeomorphisms. However, how does one compute it in general,
say for the curves in \ref{hopf}c.
There is a procedure for these cases. One should
traverse one of the curves and add 1/2 for every crossing of the type
shown in figure \ref{cross}a and -1/2 for the type \ref{cross}b.
This in particular shows
\begin{figure}[h]
\vskip 3cm
\caption{Upper and under crossings for the definition of the Gauss Linking
number}
\label{cross}
\end{figure}
that the curves in figure \ref{hopf}c
have vanishing linking number in spite of
being obviously linked (this is called the Whitehead link). This clearly
shows that we will need other, more complicated, invariants to be able
to classify linkings.

It should be apparent to the reader that these concepts are both
elegant and clearly related to the issues of Quantum Gravity we
discussed in the previous chapter. However, they do not seem quite
geared to a direct application.  Say we wanted to consider the Gauss
linking number as a candidate for a wavefunction of the gravitational
field in the loop representation (this actually fails since it is not
consistent with the Mandelstam identity).  We meet an important
requirement in the fact that it is diffeomorphism invariant. But in
the form we have casted it up to now, we cannot compute much further.
For instance, we do not know how to take an area derivative of this
quantity\footnote{Actually another technical problem appears here, in
the fact that the area derivative of diffeomorphism invariant
quantities is rather ill defined, being on a similar standing to the
derivative of a Heaviside function, since diffeomorphism invariant
functions do not change smoothly under addition of an infinitesimal
loop}.

It would be convenient to have an analytic expression for the knot invariants.
For the Gauss Linking number there actually is one. It is easy to see (for
a demonstration see Maxwell's 1873 Treatise on Electricity and Magnetism!
Volume II page 419-423) that the following expression,
\begin{equation}
{\rm GL}(\gamma_1,\gamma_2) = {\textstyle {1\over 4 \pi}}
\oint_{\gamma_1} ds \oint_{\gamma_2} dt \
\dot{\gamma_1}^a(s) \dot{\gamma_2}^b(t) \epsilon_{abc} {(\gamma_1^c(s) -
\gamma_2^c(t)) \over |\gamma_1(s) -\gamma_2(t)|^3}
\end{equation}
gives an analytic expression for the Linking number.  A very exciting
fact of the recent work on Chern-Simons theories is that it has  provided
similar analytic expressions for several other knot invariants.

We can actually write the above invariant in terms of the loop
coordinates we introduced in the last chapter. The expression is,
\begin{equation}
{\rm GL}(\gamma_1,\gamma_2) = {\textstyle {1\over 4 \pi}} X^{ax}(\gamma_1)
X^{by}(\gamma_2) \epsilon_{abc} {(x^c -y^c)\over |x-y|^3}
\end{equation}
where we again assume that repeated $x,y$ indices are integrated over the
whole manifold. In spite of its appearance (it involves a local chart of
coordinates and even a distance!) this expression is actually diffeomorphism
invariant. The fact that we can express it in terms of the loop coordinates
enormously
facilitates the application of the constraints to these expressions.
For instance, one can easily prove that the diffeomorphism constraint actually
annihilates ${\rm GL}(\gamma_1,\gamma_2)$, in accordance with the fact
that this expression is diffeomorphism invariant. (We will see soon what
happens when one applies the Hamiltonian constraint).

There are many other things to be said about knot theory. We will stop here
for reasons of space. The reader is encouraged to enjoy the books by Lou
Kauffman \cite{Ka1,Ka2}
on the subject for many other amusing and important properties of
knots.

\subsection{Knot Polynomials}

A very important step towards the construction of knot invariants, and
indirectly towards the classification problem, has been the invention of
the knot polynomials. These are polynomials in an arbitrary variable, usually
called $t$, uniquely associated with a given knot. For each knot there is
a given finite order polynomial, although the order may be different
for another knot. The important point is that the polynomials are invariant
under diffeomorphisms, therefore each coefficient is a knot invariant.
Examples of such polynomials are those of Alexander-Conway, The Kauffman
Bracket, Jones
and HOMFLY. The work on Chern-Simons theories has given rise to even newer
polynomials \cite{Guuniv}
and to previously unknown relationships among the known ones.

Knot polynomials are usually prescribed by a set of implicit relations, known
as skein relations. These relations are enough to find out the particular
polynomial associated with a given knot (sometimes the task is not totally
trivial!). Let us see an example. Consider the Conway Polynomial
$C(\gamma)[t]$, defined by the Skein Relation that appears in figure
\ref{skein}.

\begin{figure}[h]
\vskip 3cm
\caption{Skein relation for the Conway Polynomial}
\label{skein}
\end{figure}

This relation should be interpreted in the following way: for a given knot,
project it on a plane and focus on a single crossing. Cut out the crossing
and leave four incoming threads. The value of the
polynomial for the knot with the crossing drawn glued into the threads left
out is related to that of the polynomial with the other crossings drawn via
the skein relation.  This plus the fact that the polynomial on the unknot is
normalized to one is enough for computing the polynomial for an arbitrary
knot.

Let us work out an example, evaluating the Conway
Polynomial for the trefoil knot given in figure \ref{knot0}.
\begin{figure}
\vspace{3truecm}
\caption{The Trefoil knot}
\label{knot0}
\vspace{19truecm}
\caption{Evaluation of the Conway Polynomial for the trefoil knot}
\label{knot1}
\end{figure}
We focus on the crossing encircled by the dotted line and apply the
Skein Relation of figure \ref{skein}.  We see in figure
\ref{knot1} that it relates the value of the polynomial evaluated on
the knot of interest (our final result) to the value of the polynomial
for two other knots. For the left one, one can immediately see that
since the knot is homotopic to the unknot, the value of the polynomial
on it is 1.  For the link on the right we need to do some more work.
We again apply the skein relation focusing on the encircled crossing,
to evaluate the value of the polynomial on the link of interest.
Again we see in figure \ref{knot1} that one of the polynomials is
one and the other is evaluated on two unlinked curves.  Applying again
the skein relation we can see that the value of the polynomial on such
a link is vanishing, Therefore, substituting back, one sees that the
value of the Conway Polynomial for the knot of interest is $1+t^2$.

So we see that the innocent looking skein relations actually contain all the
information needed to associate a series of knot invariants to a knot (the
reader can verify that by deforming the knot in question and applying the
skein relations, one gets the same result).

Again, having a knot polynomial written as a skein relation is not very
useful for our purposes. We  would like to have a more analytic kind
of expression for the knot polynomials to, say, apply the Hamiltonian
constraint of Quantum Gravity to it and see if it is a quantum state of the
gravitational field. We will see in the
following subsections that Chern-Simons
theory has been extremely useful to find such analytic expressions for the
knot polynomials.

\subsection{Chern-Simons theory and knot polynomials}

As the reader must have noticed from previous sections, Chern-Simons theories
seem to play a crucial role concerning Knots. Dozens of articles have been
written studying various aspects of these theories and this brief subsection
can certainly not be anything else but a gross oversimplification. The
reader interested in gaining a good understanding is encouraged to explore
the relevant papers.

In a nutshell, Chern-Simons theories are gauge theories
defined in 2+1 dimensions having as action,
\begin{equation}
S_{CS}= k \int d^3 x\ {\rm Tr}( A_a \partial_b A_c +
{\textstyle {2\over 3}} A_a A_b A_c)
\epsilon^{abc}.
\end{equation}
where the connection A takes value on some group, let us fix for our
interests $SU(2)$. $k$ is the coupling constant of the theory.
Notice that no use of a spacetime metric was needed to write this action (as
one needs, for instance to write the Yang Mills action when one raises or
lowers indexes in $F_{ab} F^{ab}$). This is therefore a topological field
theory (although what is meant by this is quite context dependent). It is
invariant under diffeomorphisms. The classical equations of motion for these
theories simply state that the theory is $SU(2)$ invariant
and that the connection is
flat ($F_{ab}$ constructed out of $A_a$ is zero).
Wilson Loops are actually observables in the theory. Since the connection is
flat, when one deforms the loop with a diffeomorphism, the value of the
Wilson loop does not change. Notice that this does not happen for gravity.

Since the Wilson loop is an observable, one can ask what is its quantum
expectation value. In the language of Path Integrals,
\begin{equation}\label{expectation}
<W(\gamma)> = \int d A\ e^{i S_{CS}}\ W(A,\gamma)
\end{equation}
Now this expectation value should be invariant under diffeomorphisms, i.e.
it should be a knot invariant, parametrized by the coupling constant $k$.
In the following subsubsections we will explore in some detail what sort
of knot invariant this quantity is. We will first derive a skein relation
for it and then, taking advantage of the fact that perturbation theory
can be used to evaluate (\ref{expectation}) (remember this is Chern-Simons
theory, not Quantum Gravity!) we will present an analytic expression for
the knot invariant.

\subsubsection{A skein relation for $<W(\gamma)>$}

It turns out that one can prove that
$<W(\gamma)>$ satisfies the Skein relations of the knot polynomial known
as Kauffmann Bracket, which is intimately related to the Jones Polynomial.
This
was an important discovery due to Witten \cite{Wi}. Witten obtained his
result nonperturbatively by recurring to conformal field theories. It
turns out that the result can be obtained also in a perturbative fashion
with a more modest machinery \cite{Sm,CoGuMaMi}. The result also holds
when the loops have intersections \cite{BrGaPunpb}. We here sketch part
of the perturbative proof in order to give the reader a flavor of these
calculations and also to illustrate the usefulness in this context of
some of the techniques we introduced in section 5.

\begin{figure}[h]
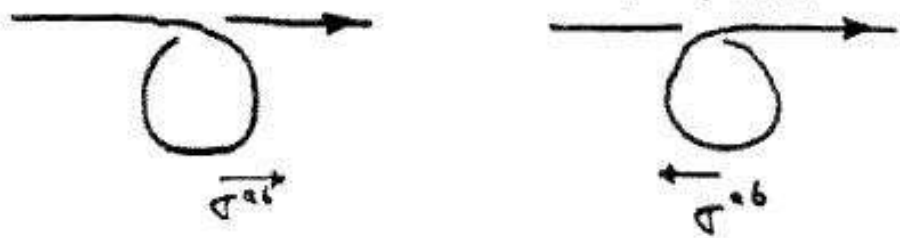

\vskip 5 cm
\caption{Under and upper crossings created by adding an oriented small loop}
\label{upund}
\end{figure}

In order to establish a skein relation for the expectation value of the
Wilson loop we basically need to relate an under and an upper crossing.
This can be done since, as illustrated in figure \ref{upund}, one can
obtain  an upper (or under, depending on the orientation of the small
loop) crossing by adding a small loop at a
given point of the loop.
It turns out that we have already developed a technique for evaluating the
change in a function of a loop when one adds a small loop: it is the loop
derivative. Therefore what we basically need to compute is the loop
derivative of $<W(\gamma)>$.

The change in $<W(\gamma)>$ due to the addition of a small loop
 of
area $\sigma^{ab}$ is given by formula (\ref{loopd}). When combined with
formula (\ref{loopdf}), it gives rise to,
\begin{equation}
\sigma^{ab} \Delta_{ab}(x) \Psi[\gamma] = \int dA \ \sigma^{ab}
F_{ab}^i(x)\ {\rm Tr}[\tau^i U(\gamma_x^x)]\ {\rm exp}(S_{CS}).
\end{equation}
Using the relation (\ref{functdoncs}), integrating by parts,
and applying (\ref{functdonw})
one obtains:
\begin{equation}
2 k \int dA \ \sigma^{ab} \epsilon_{abc} \int dy^c\ \delta(x-y)
{\rm Tr}[ \tau^i U(\gamma_x^y) \tau^i U(\gamma_y^x)]\ {\rm exp}(S_{CS})
\end{equation}
The integral depends on the volume factor
\begin{equation}
\sigma^{ab} \epsilon_{abc} dy^c \delta(x-y)
\end{equation}
which depending on the relative
orientation of the two-surface $\sigma^{ab}$ and the differential
$dy^c$ (which is tangent to $\gamma$), can lead to  $\pm 1$ or zero.
(This expression should really be regularized. We have absorbed
appropriate extra factors in the definition of the coupling c
constant so to normalize the volume to $\pm 1$).
Consequently, depending on the value of the volume there are  three
possibilities:
\begin{eqnarray}
\delta \Psi[\gamma] &=&0\\
\delta \Psi[\gamma] &=& \pm 2 k \Psi[\gamma]
\end{eqnarray}
These equations can be diagrammatically interpreted in the following
way:
\begin{equation}
\Psi[\hat{L}_\pm]-\Psi[\hat{L}_0] = \pm 2 k
\Psi[\hat{L}_0]
\end{equation}
and coincide with part of the skein relations of a known knot polynomial,
the Kauffman Bracket \cite{Ka1,Ka2}. So we see that $<W(\gamma)>$ is
actually an analytic expression for the Kauffman Bracket in the variable
$k$.

It is interesting to notice how helpful, in order to perform this
calculations, were the notion of an area derivative and of its properties.
The original derivations \cite{CoGuMaMi} did not use these concepts (although
the treatment is fully equivalent) and therefore the proof we give here
is much more economical. This is a good example where techniques developed
for gravity are of use in a particle physics problem. We will see more of
this happening in the next subsection.

The reader may be confused by figure \ref{upund}. In the past we have
considered these kinds of ``curls'' in knots as removable. We will see the
meaning of this in the next section.

\subsubsection{Perturbative calculation of the Kauffman Bracket}

Being Chern-Simons theory a renormalizable theory, one could compute
the expectation value of the Wilson Loop perturbatively. One gets as a
result a polynomial in the variable $k$ (the coupling constant of the
theory), which should provide analytic expressions for the
coefficients of the Kauffman Bracket. In this language therefore a
coefficient of the Kauffman Bracket becomes a sum of Feynmann diagrams
for the perturbative expansion of $<W(\gamma)>$. We will sketch here
the derivation. The reader should be aware that the proofs given here
are very schematic. They ignore, for instance, the presence of ghosts
(it can actually be seen that ghosts do not contribute to the order of
perturbation we are going to discuss). The complete treatment can be
seen in \cite{GuMaMiplb} and a rigorous mathematical derivation in
\cite{BaNa}.

In terms of the expression of the Wilson loop written as a function of the
Loop Coordinates (\ref{Wilson}), we can write for the expectation value,
\begin{eqnarray}
<W(\gamma)>= &&2 + <A_{ax} A_{by}> X^{ax\,by}(\gamma) +
<A_{ax} A_{by} A_{cz}> X^{ax\,by\,cz}(\gamma) + \nonumber \\
&&+ <A_{ax} A_{by} A_{cz} A_{dw}> X^{ax\,by\,cz\,dw}(\gamma) +\ldots
\end{eqnarray}
which corresponds to a diagrammatic expansion given in figure
\ref{diagramm},
\begin{figure}[t]
\vskip 5cm
\caption{Diagrammatic expansion of the Wilson Loop in a Chern-Simons theory}
\label{diagramm}
\end{figure}
where we have represented the propagator by a wavy line and the loop
coordinates of nth order as a circle with n insertions of propagators.

The Chern-Simons theory has a propagator given by,
\begin{equation}
g_{ax by} = k\ \epsilon_{abc} {(x-y)^c \over |x-y|^3}
\end{equation}
and a vertex (contracted with three propagators),
\begin{equation}
h_{ax by cz} = k^2 \int d^3 w\ \epsilon^{def}\ g_{ax dw}\ g_{by ew}\ g_{cz fw}
\end{equation}
One simply has to contract them with the propagators and vertices to get
the expression for the Feynmann diagram.

\begin{figure}[t]
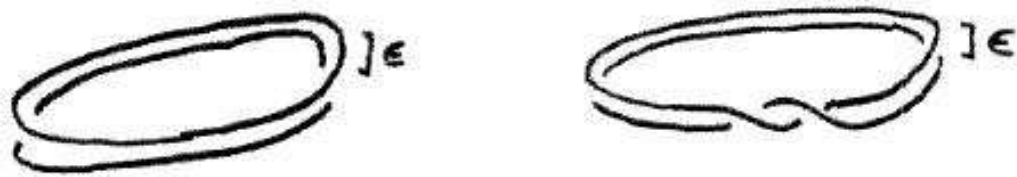

\vskip 5cm
\caption{Two different framings for a given loop. In one case the self-linking
number (computed as the linking number of the two curves defined by the
framing) is zero and in the other +1.}
\label{framing}
\end{figure}

The Feynmann diagram of first order in $k$ is therefore simply given by,
\begin{equation}
<W(A,\gamma)>^{(1)} = X^{ax by} g_{ax by}
\end{equation}
If we expand this out, remembering the expressions for $X$ and $g$ we get,
\begin{equation}
<W(A,\gamma)>^{(1)} = \oint_\gamma ds \oint_\gamma  dt \dot{\gamma}^a(s)
\dot{\gamma}^b(t) \epsilon_{abc}
{(\gamma^c(s)-\gamma^c(t)) \over |\gamma(s) -\gamma(t)|^3}
={\rm GSL}(\gamma)
\end{equation}
The reader may recognize in this expression the Gauss Linking number, except
for the fact that instead of having two knots, we have only one. We are
actually computing the linking of the knot with itself!
This sometimes is called ``Gauss self-linking number'' and we denote it as
${\rm GSL}(\gamma)$.
This presents a
small difficulty. The expression seems to be ill defined when $s=t$ since
the denominator vanishes. This is actually not true since the numerator
also vanishes, and faster. But there is a problem with this expression, if
one computes it carefully, one finds out it not only depends on the loop
but on the definition of an arbitrary normal vector to the loop
\cite{Ca,Wi,CoGuMaMi}. This means
that in the calculation diffeomorphism invariance has been broken (mildly).
A simple solution for it is
to ``frame'' the loop, i.e. convert it to a ribbon and compute the self-linking
number as the linking number of the two loops on the sides of the ribbon.
This however is not a univocous prescription since one can add twists to
the ribbon and change its value. Moreover, as we see in the figure
\ref{framing}, two equivalent loops may yield inequivalent ribbons, so
diffeomorphism invariance in the loop sense is lost when one generalizes
to ribbons.
We will see how one can deal with this
problem of the loss of diffeomorphism invariance later on.

The order $k^2$ Feynmann diagram is the sum of two terms. These terms can
be rearranged into three contributions. One of them is the square of the
self-linking number. The other two terms together give an analytic
representation for another well known knot invariant.
Details can be seen in reference \cite{CoGuMaMi}.
\begin{equation}
<W(\gamma)>^{(2)} = k^2 ({\rm GSL}(\gamma)^2+A_2(\gamma))
\end{equation}
where $A_2$ is related to the
second coefficient of the Conway Polynomial
(it is actually $(A_2 +{\textstyle
{1\over 12}})$
and is also related to the second coefficient in an expansion of the
Jones Polynomial. From now on we will loosely refer to it as ``the second
coefficient of the Jones Polynomial''. Notice that this invariant is truly
diffeomorphism invariant, i.e. it is framing-independent.

This construction goes on at higher
orders, each coefficient of the Kauffman Bracket breaks up into a framing
dependent portion function of the coefficients of lower order plus a new
knot invariant, which is framing independent. The third order expression
in $k$ is \cite{DiGaGr},
\begin{equation}
<W(\gamma)>^{(3)} = k^3 ({\rm GSL}(\gamma)^3+{\rm GSL}(\gamma) A_2(\gamma)
+A_3(\gamma)) \label{third}
\end{equation}
where $A_3(\gamma)$ is another framing independent knot invariant
related to the third coefficient in the expansion of the Jones Polynomial.
We do not have space to discuss it here, but crucial in the recombination
of the terms to forming expressions of the type (\ref{third}) is the
use of the free part of the loop coordinates, discussed in section 5.5.
If one uses the free parts of the loop coordinates it is immediate to
find expressions for the framing independent knot invariants that appear
at each order. Although this decomposition can be done ``by hand'' (and
that was the way it was done in ref. \cite{GuMaMi}), it is much more
economical to perform it using the loop coordinates. At order three and
higher, the use of loop coordinates is almost mandatory due to the
complexity of the expressions involved, and actually the first calculation
of the third order terms was done in that way. The reader is referred
to references \cite{DiGaGrLe,DiGaGr} for more details on the use of loop
coordinates to generate the expansions discussed.

Therefore,
as a consequence of this analysis we have the following analytic expression
for the Kauffman Bracket,
\begin{eqnarray}
{\rm Kauffman}\ {\rm Bracket}(\gamma)[k] =&& 1(\gamma) + {\rm GSL}(\gamma) k +
({\rm GSL}(\gamma)^2 +A_2(\gamma)) k^2 + \label{perturbative}\\
&&+({\rm GSL}(\gamma)^3 + {\rm GSL}(\gamma) A_2(\gamma) + A_3(\gamma)
) k^3+...\nonumber
\end{eqnarray}

What can we do about the framing problem?
Clearly we cannot use the Kauffman Bracket
as a candidate for a state of gravity because it is not invariant under
deformations of the loops (remember figure \ref{framing}).
However, we notice that ``buried'' inside each coefficient of the Kauffman
Bracket is present a framing independent knot invariant (a quantity that
is really invariant under deformations of the loops). What we are seeing
is the perturbative emergence of the exact relation, known
to mathematicians,
\begin{equation}
{\rm Kauffman}\ {\rm Bracket}(\gamma)[k] =e^{(k {\rm GSL}(\gamma))} {\rm
Jones} {\rm Polynomial}(\gamma)[k]
\end{equation}
so we see that all the framing dependence can be concentrated in the
``phase factor'' ${\rm exp}(k {\rm GSL}(\gamma))$. We will see that this
result allows us to construct real diffeomorphism invariant states of
gravity in the next section.

\section{Knot theory and quantum states of gravity}

The reader may seem surprised by the rather lengthy detour into
Chern-Simons theory in the last section. The reason for this will
become apparent here.

In section 4.3 we saw that there existed an exact solution to all
the constraints of Quantum Gravity in the connection representation
(with cosmological constant) given by,
\begin{equation}
\Psi_{\Lambda}^{CS}(A) = {\rm exp} (-{\textstyle{6 \over \Lambda}}
\int \tilde{\epsilon}^{abc} Tr[A_a
\partial_b A_c +{\textstyle 2 \over 3} A_a A_b A_c]).
\label{csstate}
\end{equation}

An interesting question would be: what is the counterpart of this
state in the loop representation? In general such a question goes
unanswered, since we do not how to perform the integral in the
loop transform,
\begin{equation}
\Psi_\Lambda^{CS}(\gamma) = \int ``d A'' W(A,\gamma) \Psi_\Lambda^{CS}(A).
\label{transformcs}
\end{equation}

However, the reader may notice that if one replaces in (\ref{transformcs})
the value for $\Psi_\Lambda^{CS}(A)$ given by (\ref{csstate}), one gets
back the expression for the expectation value of a Wilson loop in a
Chern Simons theory we derived last section!,
\begin{equation}
<W(\gamma)> = \int d A\ e^{i S_{CS}}\ W(A,\gamma)
\end{equation}
where the role of the coupling constant of the theory $k$ of last section
is now played by ${6 \over \Lambda}$. So we see that for the particular
wavefunction (\ref{csstate}) we can actually compute the transform into
the loop representation, and we already know the answer, it is the
Kauffman Bracket!. Again we should stress that we cannot consider this
knot polynomial strictly as a state of Quantum Gravity since it is not
diffeomorphism invariant due to the issue of framing. It is still
remarkable that we can find an analogue of state (\ref{csstate}) in the
loop representation.

If all the formalism works, the Kauffman bracket should be a solution of
the Hamiltonian constraint of Quantum Gravity with a cosmological
constant in the loop representation. Can we check this fact? We actually
can. That is what all the technology of the loop representation we developed
in section 5 is good for. We have a wavefunction in the loop representation
(the Kauffman Bracket), we can write it in terms of the loop coordinates,
(as we saw in the last section) and therefore we can apply to it the
constraints of Quantum Gravity to see if it is a solution.

Let us therefore apply the Hamiltonian constraint of Quantum Gravity
(with a cosmological constant) in
the Loop Representation $\hat{H}_\Lambda$, to the Kauffman Bracket
$\Psi_\Lambda^{CS}(\gamma)$,
\begin{equation}
\hat{H}_\Lambda \Psi_\Lambda^{CS}(\gamma) =
(\hat{H}_0 + \Lambda\ \widehat{\rm det\ q}) (1(\gamma) + \Lambda\
{\rm GSL}(\gamma) +
\Lambda^2 ({\rm GSL}(\gamma)+A_2(\gamma)) +\ldots)
\end{equation}
where $\hat{H}_0$ is the vacuum Hamiltonian constraint.

The result of this calculation is a polynomial in $\Lambda$. If it is to
vanish, it should do so order by order in $\Lambda$. This leaves us with
the following equations,
\begin{eqnarray}
\Lambda^0:&& \hat{H}_0\ 1(\gamma) = 0 \label{order0}\\
\Lambda^1:&& \widehat{\rm det\  q}\ 1(\gamma) + \hat{H}_0\
{\rm GSL}(\gamma) = 0
\label{order1}\\
\Lambda^2:&& \widehat{\rm det\ q}\ {\rm GSL}(\gamma) +
\hat{H}_0\ {\rm GSL}(\gamma)^2 \label{order2}
+\hat{H}_0\ A_2(\gamma) = 0
\end{eqnarray}
and so on for higher orders.

Equation (\ref{order0}) trivially holds, since the area derivative in the
expression of $\hat{H}_0$ (\ref{hamiltonian}) annihilates the constant.
Equation (\ref{order1}) is a bit harder to check, but one can actually
show that it holds with minor effort.

Something really interesting happens at order $\Lambda^2$, since the terms
$\widehat{\rm det\ q}\ {\rm GSL}(\gamma) +
\hat{H}_0\ {\rm GSL}(\gamma)^2$ cancel among themselves (this calculation
is rather lengthy). That means that
for the equation to hold at this order it must happen that,
\begin{equation}
\hat{H}_0\ A_2(\gamma)=0\label{eqprl}
\end{equation}
That is, the second coefficient of the Jones Polynomial has to be a solution
of the vacuum Hamiltonian constraint of Quantum Gravity! Therefore, in order
to prove that the Kauffman Bracket is a solution of the Hamiltonian
constraint with cosmological constant, it must happen that the second
coefficient of the Jones Polynomial has to be a solution of the Hamiltonian
constraint without cosmological constant.

Historically, eqs.  (\ref{order1}) and (\ref{eqprl})
were shown to hold previous
to this discovery \cite{baruch,BrGaPuprl}. Actually eq. (\ref{eqprl}) was
quite involved to prove, requiring the use of a complicated
computer algebra code.
Whereas here we find a very natural argument why it should hold.

What happens to higher orders? At each order a similar decomposition occurs
for the coefficients of the Kauffman Bracket. That led us to conjecture
\cite{BrGaPuessay} that maybe at all orders the same occurred. That is,
maybe at all orders ``nested'' inside the Kauffman Bracket was a state of
vacuum Quantum Gravity. Even better, since at each order the portion
that is a candidate to be a state of vacuum gravity is a coefficient of an
expansion of the Jones Polynomial, we could conjecture that,
\begin{equation}
\hat{H}_0\ {\rm Jones}_\Lambda(\gamma) = 0 ???
\end{equation}
Unfortunately, computations to try to prove it get more and more involved
for higher orders. There is preliminary evidence of a possible proof to
all orders involving heavy use of the loop coordinates, but we are
unprepared to report about it here \cite{DiGaGrPu}.

Notice that this state we have found for the vacuum Hamiltonian constraint
is {\em framing independent}, that is, it is a true knot invariant, and
therefore a true state of Quantum Gravity. Therefore the use of framing
dependent objects before can be seen as an intermediate artifact in the
calculation, as if one used a non-diffeomorphism invariant proof to show
that a diffeomorphism identity holds. This would be true for all the
conjectured states.

How confident can one be of this result? To put this issue in perspective,
we should list the potential points where our argumentation has been weak.
\begin{itemize}
\item Framing dependence. As we mentioned, when using the result from
Chern-Simons theory to obtain the ``loop transform'' of the Chern-Simons
state, one introduces a framing dependence. Crudely put, this means the
``loop transform'' of a wavefunction that was invariant under diffeomorphisms
fails to be invariant. We do not know how to improve this situation. The
framing dependence of the Chern-Simons result is well established and is
related to spin-statistics in three dimensions. We can just argue that
this result was used as an intermediate result and the final result, that
the Jones Polynomial coefficient solves the vacuum constraint, is framing
independent. Another possible way out of the framing difficulty would
be to abandon loops and work directly in the Coordinate Representation
mentioned at the end of section 5.6. One would then lose the connection
with knot theory but all the expressions involved could be written as
well defined functions of smooth vector densities. Unfortunately, many
details have to be worked out before we can really claim this is a
solution to the problem.
\item Regularization. The Hamiltonian constraint we are using involves a
regularization and when we claim that something is annihilated by the
constraint we really mean it is annihilated  at leading order when the
regulator is removed. A more careful study of regularization is in order.
\item The measure. When we use the Chern-Simons result for the expectation
value of the Wilson loop, implicitly we are assuming that the measure
used in Chern-Simons theory to perform the path integral is the same as
the one to be used in Quantum Gravity. This is by no means obvious. The
measure in the Loop Transform should in the end be related to the reality
conditions of the formalism and it is clear that the one in Chern-Simons
theory is, prima facie, not taking into account this fact. This is
related to the next point.
\item Reality. In Chern-Simons theory the connection is real,
whereas in Quantum Gravity it is complex. This affects our calculation of
the expectation value of the Wilson Loop. Clearly if one allows the connection
to be complex, formulae like (\ref{transform}) cease to make sense. The
integrals basically fail to converge. Even proofs like the one of the skein
relation should be taken with care in the complex case since the quantities
involved in the skein relations diverge. At the moment, lacking any control
about the reality conditions in the loop representation for Quantum Gravity,
there is very little we can say about this point. The only hope is that
the correct measure, reflecting the reality conditions, could somehow be
analytically connected with the Chern-Simons measure, and therefore the
results in Chern-Simons theory could be taken as analytic continuations to
the purely real case of the gravitational ones. It is evident that this is
just a hope and that we cannot say anything else at present.
\end{itemize}

\begin{figure}[h]
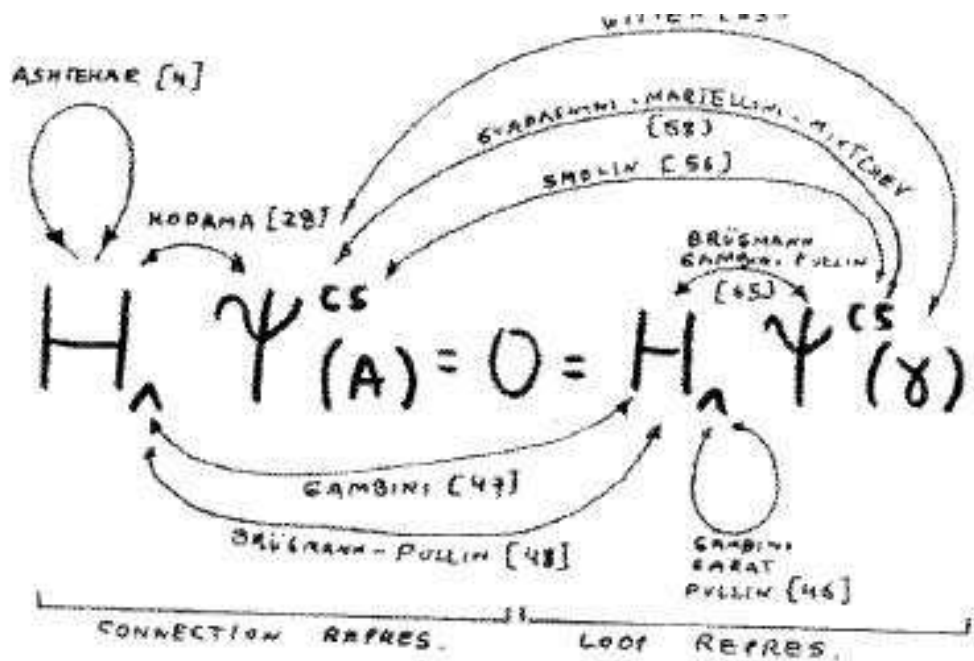

\vskip 8cm
\caption{Redundancies in the calculation offer hope that it may be correct}
\label{supportive}
\end{figure}

Given these reservations about our result, do we have any hope that it is
correct? We believe there are some supportive elements, that although far
from offering a proof, give some reassurance that our result may hold.
They are schematically shown in figure \ref{supportive}, and they can
be summarized as follows.
\begin{itemize}
\item Constraints in connection representation. These constraints were shown
to generate the correct diffeomorphism symmetry of the theory \cite{BrGaPunpb}
and to formally close the commutator algebra \cite{As}.
\item Equivalence between constraints in both representations. It was shown
both using the transform \cite{Ga} and based on the T operators of Rovelli
and Smolin \cite{BrPushr}.
\item Equivalence between wavefunctions. It was proven using perturbative
techniques in loop space \cite{CoGuMaMi,Sm} even for the intersecting case
\cite{BrGaPunpb}. It was also proven nonperturbatively \cite{Wi} and
using Feynmann diagrammatics \cite{GuMaMiplb}.
\item Constraints in the Loop Representation. Their consistency has been
partially proven at the formal level and studies are been done taking into
account regularization \cite{GaGaPu}.
\end{itemize}

All this means that if the result is wrong one or more of the previous
results should also be wrong. This could well be. For instance,
we are implicitly using
the same measure to perform the transform of the constraints and of the
states. However, even if the result is wrong, one would learn an important
lesson about various aspects of the formalism.

This was the main result we wanted to highlight in these lectures, that
using this new formalism for Canonical Quantum Gravity one could find for
the first time some nondegenerate physical states of the theory, maybe
an infinite family of them. Moreover a new branch of mathematics has been
brought into contact with Quantum Gravity, Knot Theory, both at a kinematical
level as was emphasized by Rovelli and Smolin\cite{RoSmprl} but now also
at a dynamical level, due to the role of the Jones Polynomial as a state.
It is a remarkable fact that there is a connection between General Relativity
and Knot Theory at a dynamical level. After all, the Jones Polynomial was
developed without taking into account at all the Einstein Equations. This
may just be a coincidence or it may mean that the notions of Knot Theory
are deeply intertwined with gravity in a way we do not know at present.
Will this mean that the Jones Polynomial is a state of {\em any} theory
of gravity one proposes? At present we can just offer this as a conjecture.

Assuming the Jones Polynomial is a state, as conjectured, how general a state
can it be? Mathematicians seem to agree that the Jones Polynomial is not
enough to solve the problem of knot theory, classify inequivalent knots.
That means other invariants are to be found in the future that are more
powerful. In this view, one would also expect to find states of Quantum
Gravity among them, and therefore the conjectured present family of states
would be incomplete. A recent trend in mathematics is to consider Vassiliev
invariants as more general invariants to classify knots. It is remarkable
that these invariants are defined for loops with intersections, exactly the
kind of loops that are relevant for gravity. It seems that our generalization
of the Jones Polynomial for loops with intersections \cite{BrGaPunpb} is
{\em not} a Vassiliev invariant. However, a more careful study of this
aspect is in order.
\section{Final Remarks}

Due to space limitations these lecture notes can only be taken as a ``tourist
brochure'' of the subject in question. Many oversimplifications have been
introduced that allow the reader to quickly view several important
results, but may also obscure a detailed understanding of the topics. We urge
the readers who want more than a lax overview to consult the appropriate
references. We acknowledge that chronologically this may be the first
complete account of these findings that sees the light through publication.
We urge the interested readers to pay attention, since in the immediate future
more detailed accounts of these topics will be published.
We would like
to finish by referring to some topics that were not even discussed in the
text and making some final remarks on the present status and prospect of
the subject. Even here we will have to be unfair and leave unmentioned
important topics.

In these lectures we have reviewed basically three things.
\begin{itemize}
\item The Ashtekar reformulation of General Relativity.
\item Some attempts to construct canonical quantizations using these variables.
\item The relation of some results from Chern-Simons theories to the Loop
Representation of Quantum Gravity.
\end{itemize}
The first item is clearly of great importance. The Ashtekar variables are
finding new applications in Classical Relativity every day and will certainly
become a standard tool of analysis for Relativists {\em even if attempts to
quantize the theory using them fail}. Of the many aspects not even mentioned
in these notes concerning this subject, we would like to point out to the
reader the following: a) The Capovilla-Dell-Jacobson Lagrangian reformulation
of the theory {\em purely in terms of a connection} \cite{CaDeJa}.
This was a long-cherished
dream of many relativists. The work also presents a novel way of solving the
constraints which may have implications for the issue of free data of the
theory. b) The work of Samuel \cite{Sa} and Torre \cite{To}
that showed how instantons can be
transferred from Yang-Mills theory to General Relativity and their stability
analysis showing they can actually be a countable number in some cases. c)
The application of the variables to Bianchi cosmologies \cite{Ko,Korev,AsPu},
offering a new
picture of the classical (and maybe quantum) dynamics of these systems. d)
The Newman-Rovelli \cite{NeRo}
method for solving the constraint equations using a
Hamilton-Jacobi reformulation. d) The possibility that topology change
\cite{Ho} and
negative energy \cite{Va}
may occur in the theory. Summarizing, this is a healthy
area of research in which many new and important developments will be studied.

In the second item we count the connection and loop representations. These
may or may not succeed in providing a basis for a quantum theory of gravity.
Even in the case of failure, it is clear that many lessons have been learnt
from their use. We list some important pieces of work
on these areas not mentioned
in these notes a) The application to two \cite{HuSm}
and one \cite{HuPu} killing vector spacetimes,
allowing in some cases to find observables in the systems. Recent work also
shows that the quantization scheme for one polarization two Killing vector
fields may coincide with the usual quantization based on the equivalence
with a scalar field \cite{AsVa}.
b) The work on 2+1 gravity, which shows for a model
system how the connection and loop representations and the loop transform
can actually be given a rigorous meaning \cite{AsHuRoSaSm,Asstri}.
b) The application of loop
techniques to Gauge theories on the lattice \cite{Ar,Br}, linearized gravity
\cite{AsRoSmli} and  Maxwell theory \cite{AsRo}, again offering
test cases where the quantization program works to the end. c) The work on
C-P violation \cite{AsBaJo}.
d) The discussion of the issue of time in the linearized theory \cite{Asog}.

An aspect that cannot be overstressed is the development in the loop
representation, of techniques for writing differential operators in loop
space and to write wavefunctions and knot invariants in analytic form
\cite{DiGaGrLe}. These
findings transcend the area of Quantum Gravity and have immediate application
in the quantization of gauge theories (in the continuum and lattice). In
fact, we have seen some examples of their application in the sections on
Chern-Simons theory. They can also become standard tools of analysis for
knot theorists. In fact, understanding in this area is just beginning and
we may see even more progress in the near future. Of particular interest
is the Coordinate Representation of section 5.2 that could allow for the
quantization of diffeomorphism invariant theories without the
framing ambiguities of the Loop Representation.

On the final item, we can just say that it is work in progress and that in the
end technical difficulties may hamper further development or even disprove
the present results. An interesting
point seems to be that some of the results seem to survive the inclusion
of matter \cite{GaPu}.
Another result is that some notions of Knot theory seem to be
useful to select an inner product for the theory at least in some toy
subsectors \cite{Ba}.
Moreover, the first hint of what a semiclassical interpretation
may look like in this context is starting to emerge \cite{AsRoSmwe}.

Notice that our treatment has evaded the ``big questions'' of Quantum
Gravity, as what is the inner product, the issue of observables and
the issue of time. The only comment we can make is that the fact of
being able to explore (tentatively) the space of states of the theory
may provide a better framework in which to address these problems in
the future.

We do not know what will the outcome of this --at the moment--
happy marriage
of Knot Theory and Quantum Gravity be. As in any other case where a new
mathematical technique is introduced into an area of Physics there is
potential for striking new results and also for a lot of red herrings.
Only time and much more effort will decide which of these two
situations  we are actually creating with our work.

\section*{Acknowledgments}

I am most grateful to the organizers, especially Jose Luis Lucio and
Octavio Obreg\'on for giving me the opportunity to
present my views on these topics. I am also grateful to the faithful audience
that endured my lectures in Guanajuato.
Most of the original research presented in sections 4 and subsequents
was done in
collaboration with Rodolfo Gambini (Montevideo) and Bernd Br\"ugmann
(Syracuse). Collaborating with them has been a great pleasure and has
enriched a lot my view of the subject. I am also indebted by their
careful reading of this manuscript. My understanding (or lack thereof) of
the New Variables
and the quantization program have been shaped to a great extent through
interactions with
Abhay Ashtekar,
John Baez,
Daniel Boyanovsky,
Louis Crane,
Mario D\'{\i}az,
Cayetano Di Bartolo,
Alcides Garat,
Gabriela Gonz\'alez,
Jorge Griego,
Viqar Husain,
Ted Jacobson,
Lou Kauffman,
Karel Kucha\v{r},
Renate Loll,
Pablo Mora,
Ted Newman,
Richard Price,
Joe Romano,
Carlo Rovelli,
Joseph Samuel,
Lee Smolin,
Ranjeet Tate,
Peter Thomi,
Paul Tod,
Charles Torre,
Claes Uggla,
Madhavan Varadarajan,
Enric Verdaguer and many other visitors to the Syracuse and Utah groups.
To all of them I am indebted for
the insights they offered on various topics.
This work was supported in part by grant NSF PHY92-07225 and by research
funds of the University of Utah.

\end{document}